# Quantum Computer: Hello, Music!

Eduardo Reck Miranda
Interdisciplinary Centre for Computer Music Research (ICCMR)
University of Plymouth, UK
eduardo.miranda@plymouth.ac.uk

## 1 Introduction

Quantum computing is emerging as a promising technology, which is built on the principles of subatomic physics. By the time of writing, fully fledged practical quantum computers are not widely available. But research and development are advancing at exponential speeds. Various software simulators are already available [1, 2]. And a few companies have already started to provide access to quantum hardware via the cloud [3, 4]. These initiatives have enabled experiments with quantum computing to tackle some realistic problems in science; e.g., in chemistry [5] and cryptography [6].

In spite of continuing progress in developing increasingly more sophisticated hardware and software, research in quantum computing has been focusing primarily on developing scientific applications. Up till now there has been virtually no research activity aimed at widening the range of applications of this technology beyond science and engineering. In particular applications for the entertainment industry and creative economies.

We are championing a new field of research, which we refer to as *Quantum Computer Music*. The research is aimed at the development of quantum computing tools and approaches to creating, performing, listening to and distributing music.

This chapter begins with a brief historical background. Then, it introduces the notion of algorithmic music and presents two quantum computer music systems of our own design: a singing voice synthesizer and a musical sequencer. A primer on quantum computing is also given. The chapter ends with a concluding discussion and advice for further work to develop this new exciting area of research.

## 2 Historical Background

As early as the 1840s, mathematician and allegedly the first ever software programmer, Lady Ada Lovelace, predicted in that machines would be able to compose music. On a note about Charles Babbage's Analytical Engine, she wrote:

*"Supposing, for instance, that the fundamental relations of pitched sounds in the science of harmony and of musical composition were susceptible of such expression and adaptations, the Engine might compose elaborate and scientific pieces of music of any degree of complexity or extent."* ([7], p.21)

People hardly ever realize that musicians started experimenting with computing far before the emergence of the vast majority of scientific, industrial and commercial computing applications in existence today. For instance, in the 1940s, researchers at Australia's Council for Scientific and Industrial Research (CSIR) installed a loudspeaker on their Mk1 computer (Figure 1) to track the progress of a program using sound. Subsequently, Geoff Hill, a mathematician with a musical background, programmed this machine to playback a tune in 1951 [8].

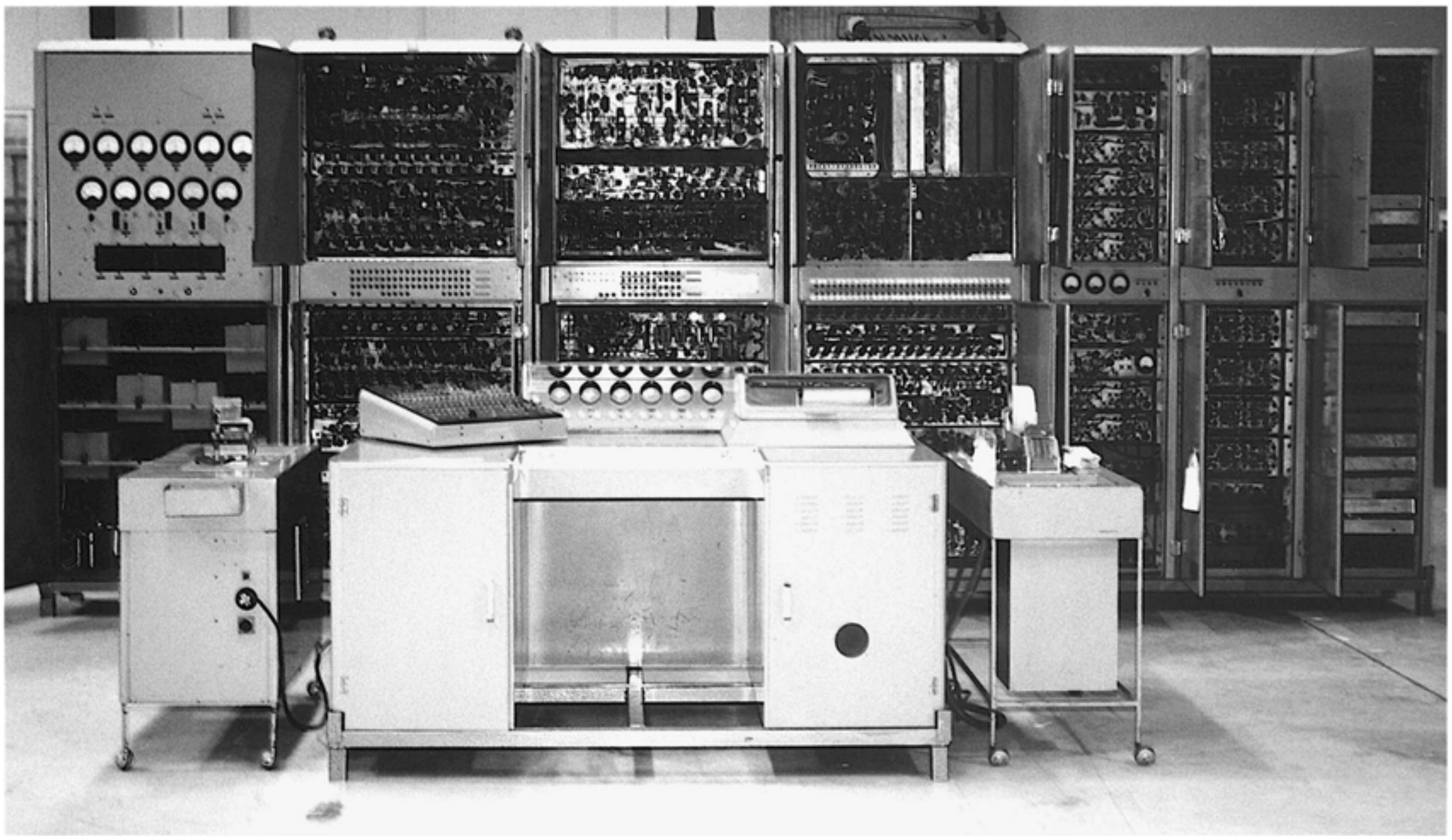

**Figure 1:** CSIRAC computer used to playback a tune in the early 1950s. The loudspeaker can be seen in the right-hand door of the console. (Image published with the kind permission of Prof Paul Doornbusch.)

And in the 1950s composer and Professor of Chemistry, Lejaren Hiller collaborated with mathematician Leonard Isaacson, Lejaren Hiller and Leonard Isaacson, at University of Illinois at Urbana-Champaign, programmed the ILLIAC computer to compose a string quartet entitled *Illiac Suit*e. The ILLIAC, short for Illinois Automatic Computer, was one of the first mainframe computers built in the USA, comprising thousands of vacuum tubes. The *Illiac Suit*e consists of four movements, each of which using different methods for generating musical sequences, including hard-coded rules and a probabilistic Markov chain method [9]. This string quartet is often cited as a pioneering piece of algorithmic computer music. That is, whereas Mk1 merely played back an encoded tune, ILLIAC was programmed with algorithms to compose music.

Universities and companies have been welcoming musicians to join their research laboratories ever since. A notable early example is AT&T's Bell Laboratories, in New Jersey, where in the early 1960s composer Max Mathews developed MUSIC III: a system for synthesizing sounds on the IBM 7094 computer.  Descendants of MUSIC III are still used today; e.g., programming languages for audio such as Csound [10].

The great majority of computer music pioneers were composers interested in inventing new music and/or innovative approaches to compose. They focused on developing algorithms to generate music. Hence the term 'algorithmic music'. In addition to those innovators cited above, names such Iannis Xenakis, Pietro Grossi, Jean-Claude Risset and Charles Dodge, amongst a few others, come to mind. Those early pioneers of Computer Music unwittingly paved the way for the development of a thriving global music industry.

Computers play a pivotal part in the music industry today. And emerging quantum computing technology will most certainly have an impact in the way in which we create and distribute music in time to come. Hence the dawn of Quantum Computer Music is a natural progression for music technology.

Prior to this chapter, the ICCMR team published preliminary studies with photonic quantum computing [11] and with Grover's search algorithm to produce melodies [12]. A book chapter about *Zeno*, a composition for bass clarinet and music generated by a quantum computer interactively, is also available [13].

## 3 Algorithmic Computer Music

The first uses of computers in music were for running algorithms to generate music. Essentially, the art of algorithmic music consists of (a) harnessing algorithms to produce patterns of data and (b) developing ways to translate these patterns into musical notes or synthesised sound. An early approach to algorithmic music, which still remains popular to date, is to program the computer to generate notes randomly and then reject those that do not satisfy given criteria, or rules. Musical rules based classic treatises on musical composition (e.g., [14]) are relatively straightforward to encode in a piece of software.

Another widely used approach employs probability distribution functions to predispose the system towards picking specific elements from a given set of musical parameters. For instance, consider the following ordered set of 8 notes, which constitute a C4 major scale: {C4, D4, E4, F4, G4, A4, B4, C5} (Figure 2). A Gaussian function could be used to bias the system to pick notes from the middle of the set. That is, it would generate sequences with higher occurrences of F4 and G4 notes.

A Gaussian function may well be viewed as a simple abstract musical rule. Abstract rules for musical composition can be expressed in a number of ways, including graphs, set algebra, Boolean expressions, finite state automata and Markov chains, to cite but a few. An example using a Markov chain to encode rules for generating sequences of notes is given below. A more detailed introduction to various classic algorithmic composition methods is available in [15].

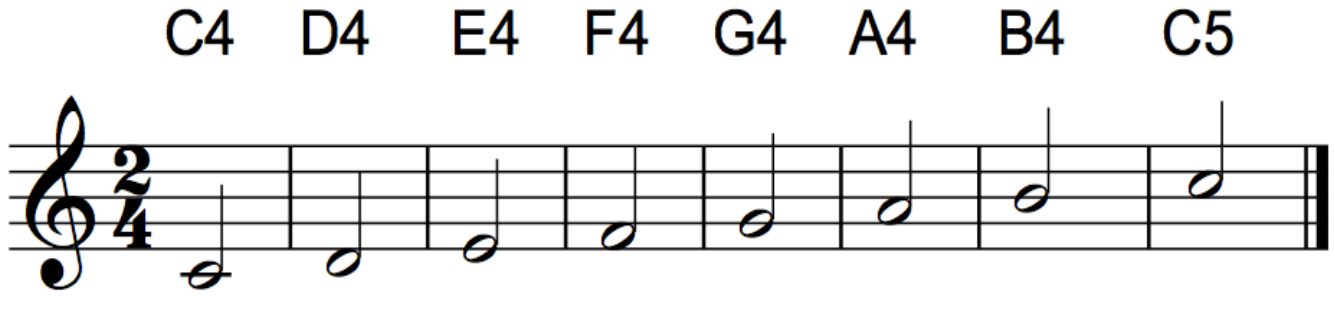

**Figure 2:** A given ordered set of musical notes.

As an example, consider the ordered set shown in Figure 2. Let us define the following sequencing rules for establishing which notes are allowed to follow a given note within the set:

Rule 1: if C4, then either C4, D4, E4, G4 or C5
Rule 2: if D4, then either C4, E4 or G4
Rule 3: if E4, then either D4 or F4
Rule 4: if F4, then either C4, E4 or G4
Rule 5: if G4, then either C5, F5, G5 or A5
Rule 6: if A4, then B4
Rule 7: if B4, then C5
Rule 8: if C5, then either A4 or B4

Each of these rules represents the transition probabilities for the next note to occur in a sequence. For example, after C4, each of the five notes C4, D4, E4, G4 and C5 has a 20% chance each of occurring.

The rules above can be expressed in terms of probability arrays. For instance, the probability array for note C4 is *p*(C4) = [0.2, 0.2, 0.2, 0.0, 0.2, 0.0, 0.0, 0.2] and for note D4 is *p*(D4) = [0.33, 0.0, 0.33, 0.0, 0.33, 0.0, 0.0, 0.0], and so on. The probability arrays for all rules can be arranged in a two-dimensional matrix, thus forming a Markov chain, as shown in Figure 3.

| | **C4** | **D4** | **E4** | **F4** | **G4** | **A4** | **B4** | **C5** |
|---|---|---|---|---|---|---|---|---|
| **C4** | 0.2 | 0.2 | 0.2 | 0.0 | 0.2 | 0.0 | 0.0 | 0.2 |
| **D4** | 0.33 | 0.0 | 0.33 | 0.0 | 0.33 | 0.0 | 0.0 | 0.0 |
| **E4** | 0.0 | 0.5 | 0.0 | 0.5 | 0.0 | 0.0 | 0.0 | 0.0 |
| **F4** | 0.33 | 0.0 | 0.33 | 0.0 | 0.33 | 0.0 | 0.0 | 0.0 |
| **G4** | 0.25 | 0.0 | 0.0 | 0.25 | 0.25 | 0.25 | 0.0 | 0.0 |
| **A4** | 0.0 | 0.0 | 0.0 | 0.0 | 0.0 | 0.0 | 1.0 | 0 |
| **B4** | 0.0 | 0.0 | 0.0 | 0.0 | 0.0 | 0.0 | 0.0 | 1.0 |
| **C5** | 0.0 | 0.0 | 0.0 | 0.0 | 0.0 | 0.5 | 0.5 | 0.0 |

**Figure 3:** Sequencing rules represented as a Markov chain.

Given a starting note, the system then picks the next based on the transition probability on the corresponding column to pick the next, and so on. An example of a melody generated using this method is shown in Figure 4.

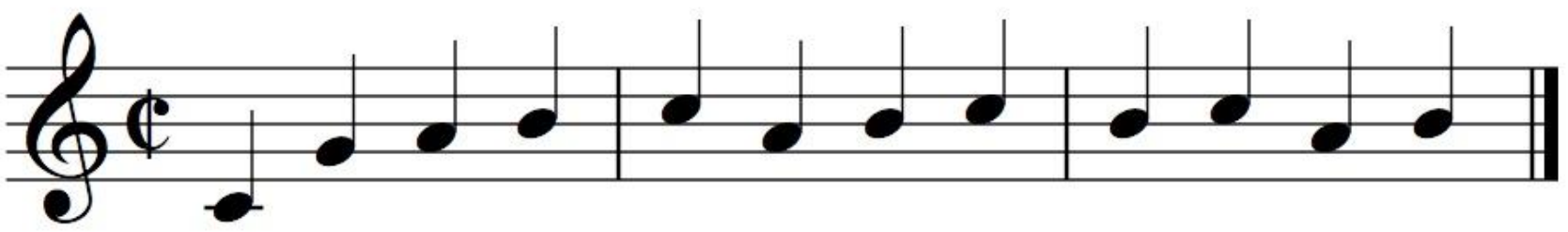

**Figure 4:** An example generated by the Markov chain shown in Figure 3.

A Markov chain whose matrix representation has non-zero entries immediately on either side of the main diagonal, and zeros everywhere else constitutes an example of a simple *random walk* process.

Imagine that a robot is programmed to play an instrument with 8 keys, to produce the notes shown in Figure 2. However, the robot is programmed in such a way that notes can be played up and down the keyboard by stepping only one key at a time. That is, only the next neighbouring key can be played. If the robot has a probability $p$ to play the key on the left side of the current key, then it will have the probability $q = 1 - p$ to go to the right. This is represented in the matrix shown in Figure 5. Random walk processes are normally represented as directed graphs, or digraphs, as shown in Figure 6.

| | **C4** | **D4** | **E4** | **F4** | **G4** | **A4** | **B4** | **C5** |
|---|---|---|---|---|---|---|---|---|
| **C4** | **0.0** | *1.0* | 0.0 | 0.0 | 0.0 | 0.0 | 0.0 | 0.0 |
| **D4** | *0.5* | **0.0** | *0.5* | 0.0 | 0.0 | 0.0 | 0.0 | 0.0 |
| **E4** | 0.0 | *0.5* | **0.0** | *0.5* | 0.0 | 0.0 | 0.0 | 0.0 |
| **F4** | 0.0 | 0.0 | *0.5* | **0.0** | *0.5* | 0.0 | 0.0 | 0.0 |
| **G4** | 0.0 | 0.0 | 0.0 | *0.5* | **0.0** | *0.5* | 0.0 | 0.0 |
| **A4** | 0.0 | 0.0 | 0.0 | 0.0 | *0.5* | **0.0** | *0.5* | 0.0 |
| **B4** | 0.0 | 0.0 | 0.0 | 0.0 | 0.0 | *0.5* | **0.0** | *0.5* |
| **C5** | 0.0 | 0.0 | 0.0 | 0.0 | 0.0 | 0.0 | *1.0* | **0.0** |

**Figure 5:** A Markov chain for random walk.

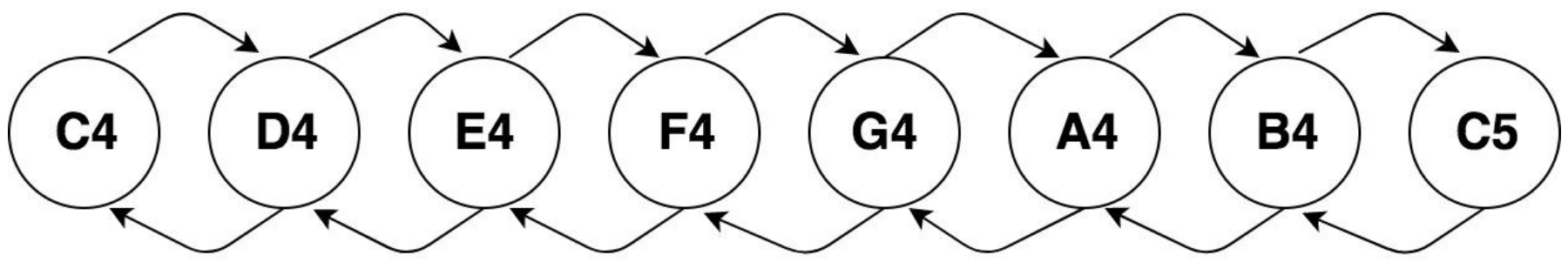

**Figure 6:** Digraph representation of the random walk scheme depicted in Figure 5.

Random walk processes are useful for generating musical sequences that require smooth gradual changes over the material rather than large jumps. Figure 7 shows an example sequence generated by our imaginary robot.

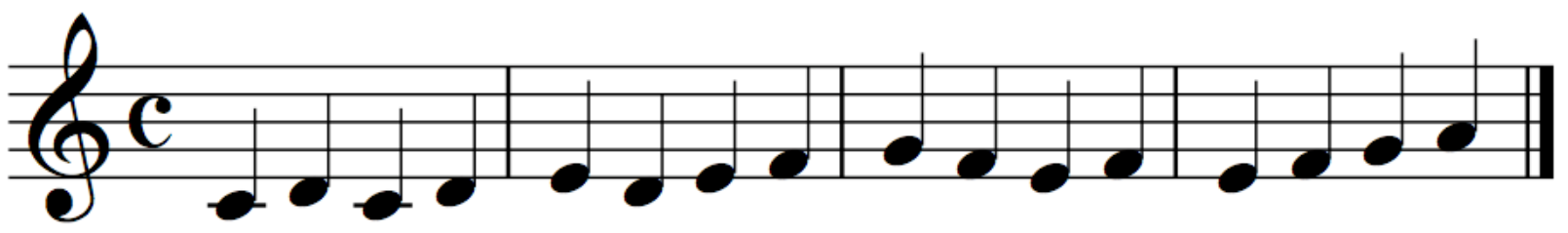

**Figure 7:** A sequence of notes generated by the random walk robot.

As computers became increasingly portable and faster, musicians started to program them to create music interactively, during a performance. Let us say, a performer plays a musical note. The computer listens to the note and produces another one as a response. Most algorithmic music methods that were developed for batch processing of music can be adapted for interactive processing. For instance, given the Markov chain above, if a performer plays the note C5, then the system would respond with A4 or B4, and so on.

A sensible approach to get started with Quantum Computer Music is to revisit existing tried-and-tested algorithmic music methods with a view to running them on quantum computers. Sooner or later new quantum-specific methods are bound to emerge from these exercises.

## 4 Quantum Computing Primer

This section provides a preliminary introduction to quantum computing, aimed at demonstrating how it differs from classical computing. It introduces the basics to follow the systems discussed in this chapter. The reader is referred to [16, 17, 18, 19] for more detailed explanations.

Classical computers manipulate information represented in terms of binary digits, each of which can value 1 or 0. They work with microprocessors made up of billions of tiny switches that are activated by electric signals. Values 1 and 0 reflect the on and off states of the switches.

In contrast, a quantum computer deals with information in terms of quantum bits, or *qubits*. Qubits operate at the subatomic level. Therefore, they are subject to the laws of quantum physics.

At the subatomic level, a quantum object does not exist in a determined state. Its state is unknown until one observes it. Before it is observed, a quantum object is said to behave like a wave. But when it is observed it becomes a particle. This phenomenon is referred to as the *wave-particle duality*.

Quantum systems are described in terms of wave functions. A wave function represents what the particle would be like when a quantum object is observed. It expresses the state of a quantum system as the sum of the possible states that it may fall into when it is observed. Each possible component of a wave function, which is also a wave, is scaled by a coefficient reflecting its relative weight. That is, some states might be more likely to be observed than others. Metaphorically, think of a quantum system as the spectrum of a musical sound, where the different amplitudes of its various wave-components give its unique timbre. As with sound waves, quantum wave-components interfere with one another, constructively and destructively. In quantum physics, the interfering waves are said to be *coherent.* As we will see later, the act of observing the wave decoheres it. Again metaphorically, it is as if when listening to a musical sound one would perceive only one of its spectral components; probably the one with the highest energy, but not necessarily so.

Qubits are special because of the wave-particle duality. Qubits can be in an indeterminate state, represented by a wave function, until they are read out. This is known as *superposition.* A good part of the art of programming a quantum computer involves manipulating qubits to perform operations while they are in such indeterminate state. This makes quantum computing fundamentally different from digital computing.

A qubit can be implemented in a number of ways. All the same, the qubits of a quantum processor need to be isolated from the environment in order to remain coherent to perform computations. The environment causes interferences that destroy coherence. One of the worst enemies of coherence is heat. A Quantum Processing Unit (QPU) has to be cooled to near absolute zero temperature to function; that is, –273.15 degrees Celsius. This is the point at which the fundamental particles of nature would stop moving due to thermal fluctuations, and retain only the so-called zero-point energy quantum mechanical motion. But even then, it is very hard to shield a QPU from the effects of our environment. In practice, interactions with the environment cannot be completely avoided, only minimized.

In order to picture a qubit, imagine a transparent sphere with opposite poles. From its centre, a vector whose length is equal to the radius of the sphere can point to anywhere on the surface. In quantum mechanics this sphere is called *Bloch sphere* and the vector is referred to as a *state vector*. The opposite poles of the sphere are denoted by $|0\rangle$ and $|1\rangle$, which is the notation used to represent quantum states (Figure 8).

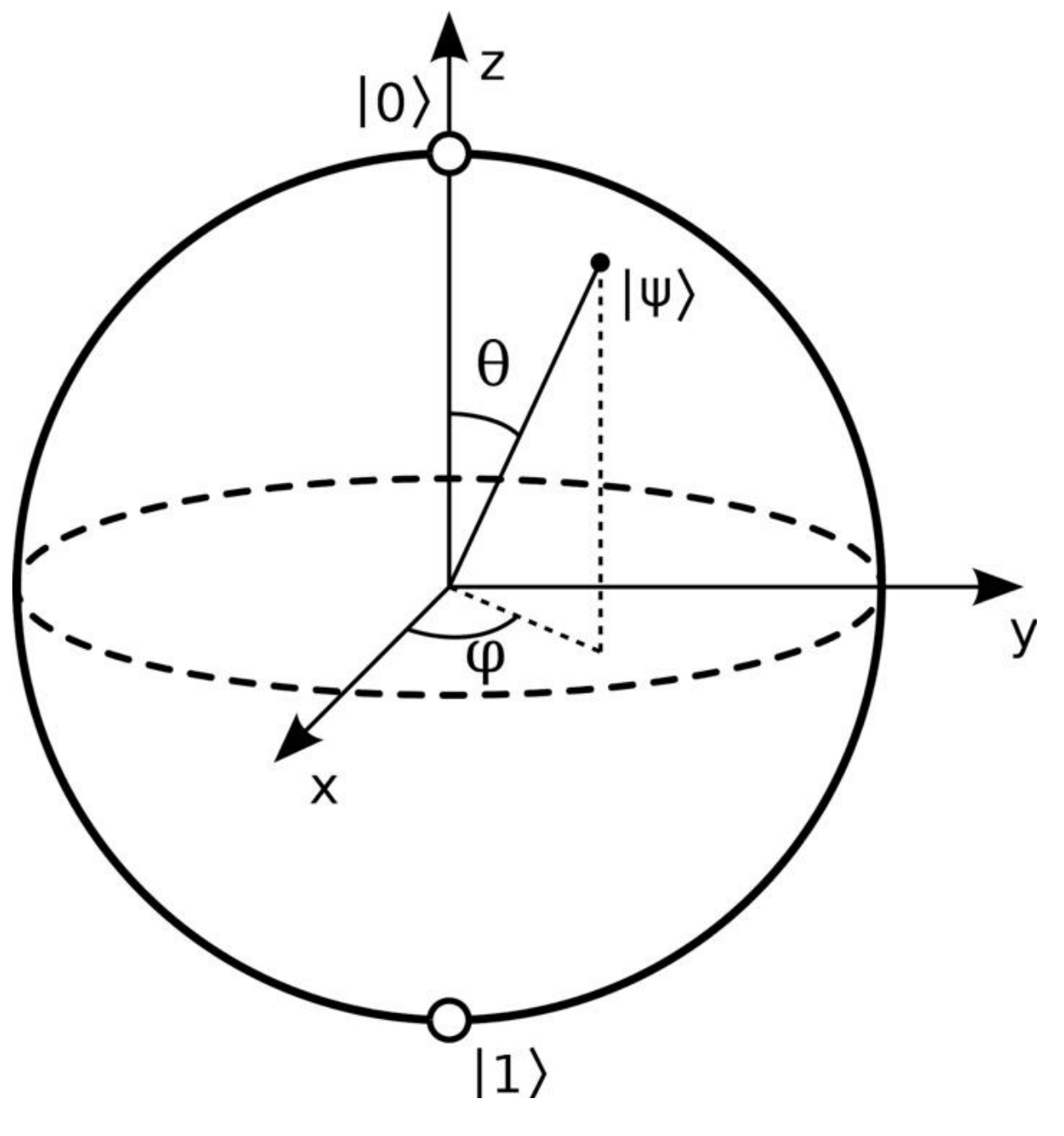

**Figure 8:** Bloch sphere. (Source: Smite-Meister, https://commons.wikimedia.org/w/index.php?curid=5829358)

A qubit's state vector can point at anywhere on the surface of the Bloch sphere. Mathematically, it is described in terms of polar coordinates using two angles, $\theta$ and $\varphi$. The angle $\theta$ is the angle between the state vector and the z-axis (latitude) and the angle $\varphi$ describes vector's position in relation to the x-axis (longitude).

It is popularly said that a qubit can value 0 and 1 at the same time, but this is not entirely accurate. When a qubit is in superposition of $|0\rangle$ and $|1\rangle$, the state vector could be pointing anywhere between the two. However, we cannot really know where exactly a state vector is pointing to until we read the qubit. In quantum computing terminology, the act of reading a qubit is referred to as 'observing' or 'measuring' it. Measuring the qubit will make the vector point to one of the poles and return either 0 or 1 as a result.

The state vector of a qubit in superposition state is described as a linear combination of two vectors, $|0\rangle$ and $|1\rangle$, as follows:

$$|\Psi\rangle = \alpha|0\rangle + \beta|1\rangle, \text{ where } |\alpha|^2 + |\beta|^2 = 1.$$

The state vector $|\Psi\rangle$ is a superposition of vectors $|0\rangle$ and $|1\rangle$ in a two-dimensional complex space, referred to as *Hilbert space*, with amplitudes $\alpha$ and $\beta$. Here the amplitudes are expressed in terms of Cartesian coordinates; but bear in mind that these coordinates can be complex numbers.

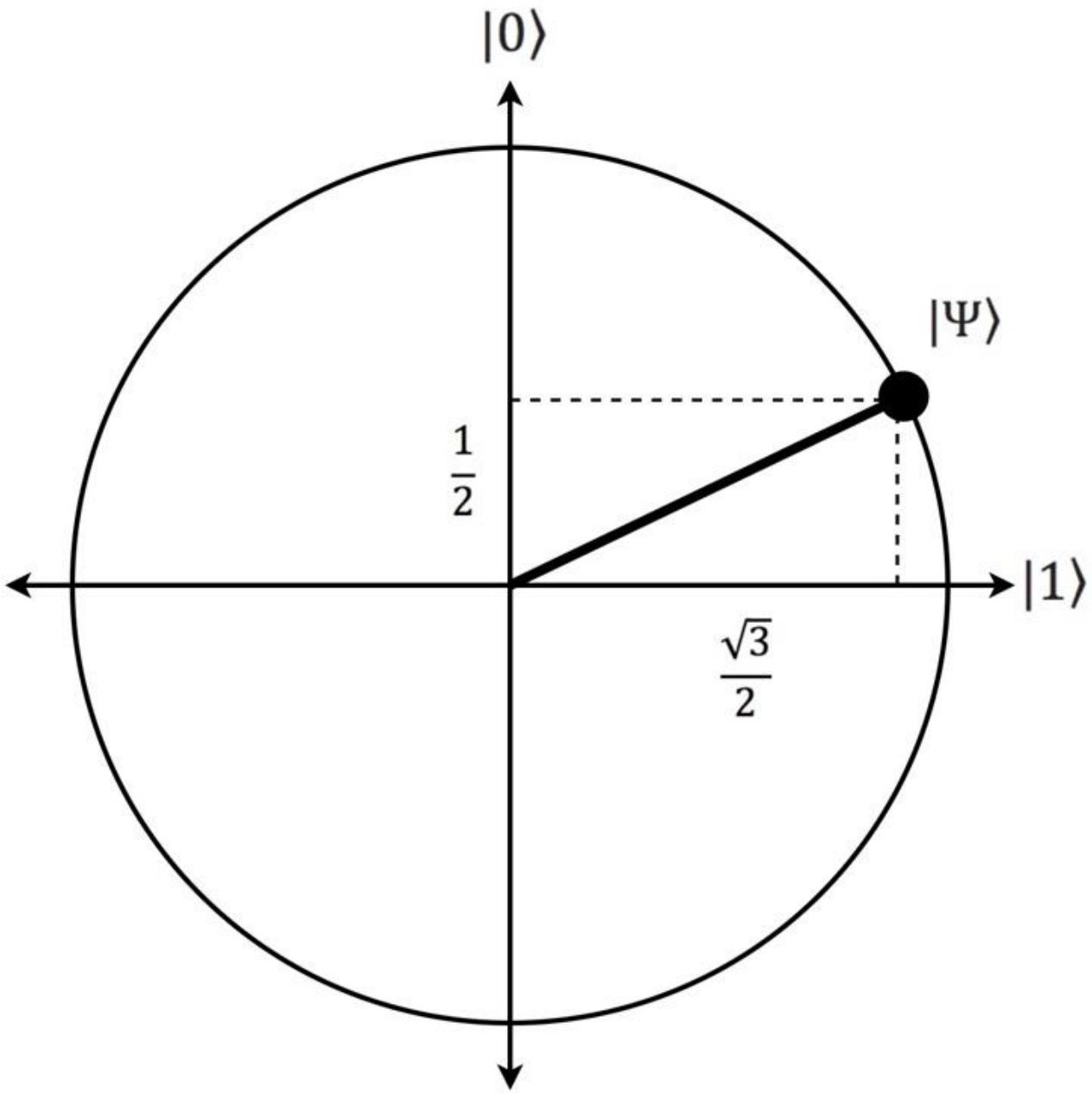

**Figure 9:** An example of superposition, where the state vector has a 25% chance of settling to $|0\rangle$ and a 75% chance of settling to $|1\rangle$ after the measurement.

In a nutshell, consider the squared values of $\alpha$ and $\beta$ as *probability values* representing the likelihood of the measurement return 0 or 1. For instance, let us assume the following:

$$|\Psi\rangle = \alpha|0\rangle + \beta|1\rangle, \text{ where } \alpha = \frac{1}{2} \text{ and } \beta = \frac{\sqrt{3}}{2}$$

In this case, $|\alpha|^2 = 0.25$ and $|\beta|^2$ = 0.75. This means that the measurement of the qubit has a 25% chance of returning 0 and a 75% chance of returning 1 (Figure 9).

Quantum computers are programmed using sequences of commands, or *quantum gates*, that act on qubits. For instance, the 'not gate', performs a rotation of 180 degrees around the x-axis. Hence this gate is often referred to as the 'X gate' (Figure 10). A more generic rotational R($\vartheta$) gate is available for quantum programming, where the angle for the rotation is specified. This gate is applicable to rotate on any of the three axes, notated as Rx, Ry, and Rz. Therefore, Rx(180) applied to $|0\rangle$ or $|1\rangle$ is equivalent to applying X to $|0\rangle$ or $|1\rangle$. In essence, all quantum gates perform rotations, which change the amplitude distribution of the system.

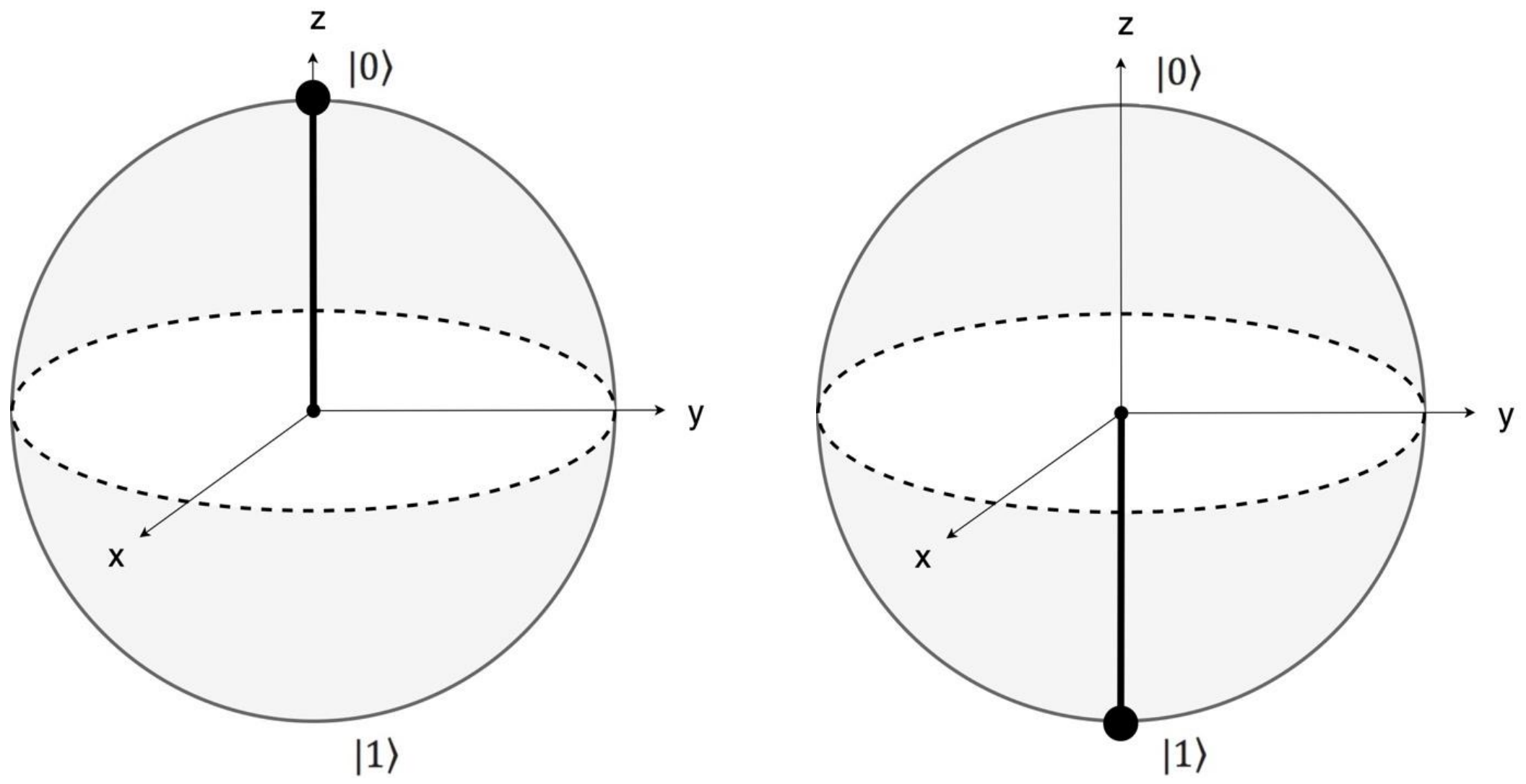

**Figure 10:** X gate rotates the state vector (pointing upwards on the figure on the left) by 180 degrees around the x-axis (pointing downwards on the figure on the right) .

An important gate in the Hadamard gate (referred to as the 'H gate'). It puts the qubit into a superposition state consisting of an equal-weighted combination of two opposing states: $|\Psi\rangle = \alpha|0\rangle + \beta|1\rangle$ where $|\alpha|^2 = 0.5$ and $|\beta|^2 = 0.5$ (Figure 11). For other gates, please consult the references given at the beginning of this section.

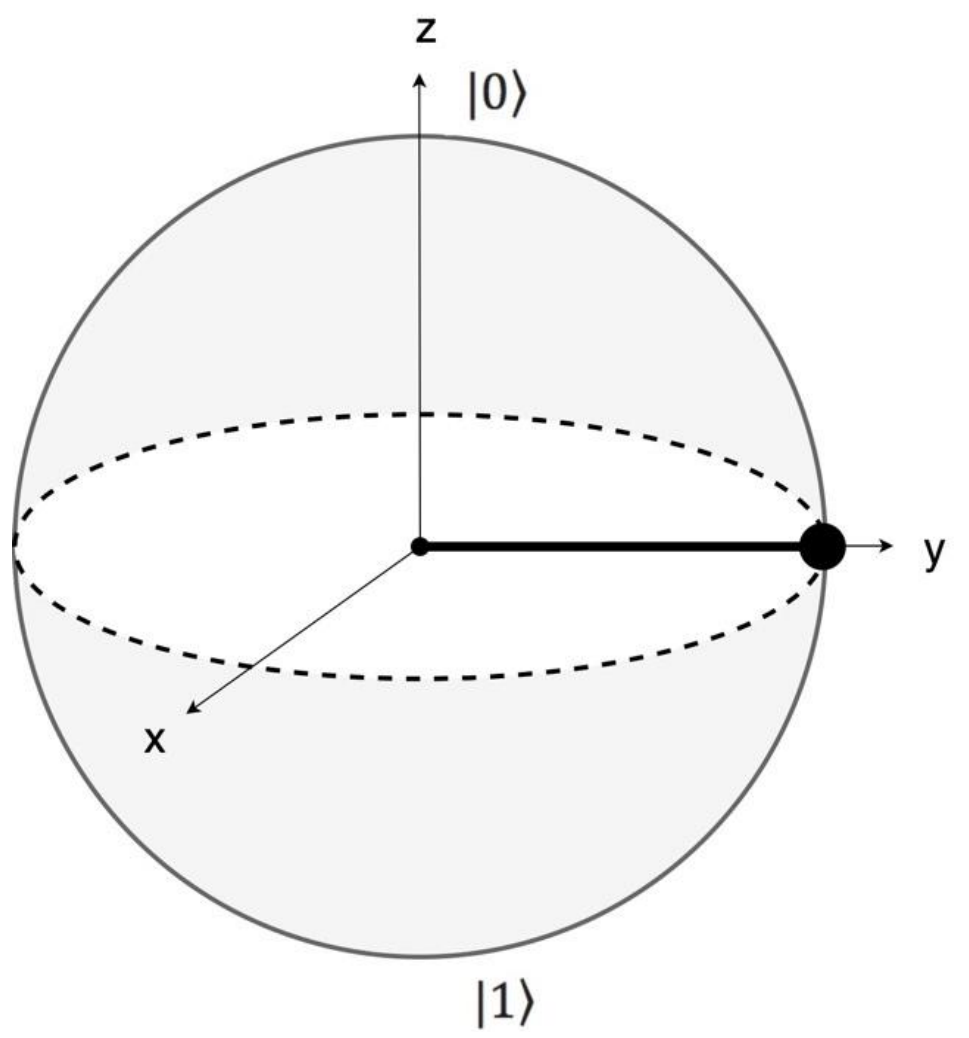

**Figure 11:** The Hadamard gate puts the qubit into a superposition state halfway two opposing poles.

Qubits typically start at $|0\rangle$ and then a sequence of gates are applied. Then, the qubits are read and the results are stored in standard digital memory, which are accessible for further handling. Normally a quantum computer works alongside a classical computer, which in effect acts as the interface between the user and the quantum machine. The classical machine enables the user to handle the measurements for practical applications; e.g., convert the measurement into musical notes.

A quantum program is often depicted as a circuit diagram of quantum gates, showing sequences of gate operations on the qubits (Figure 12).

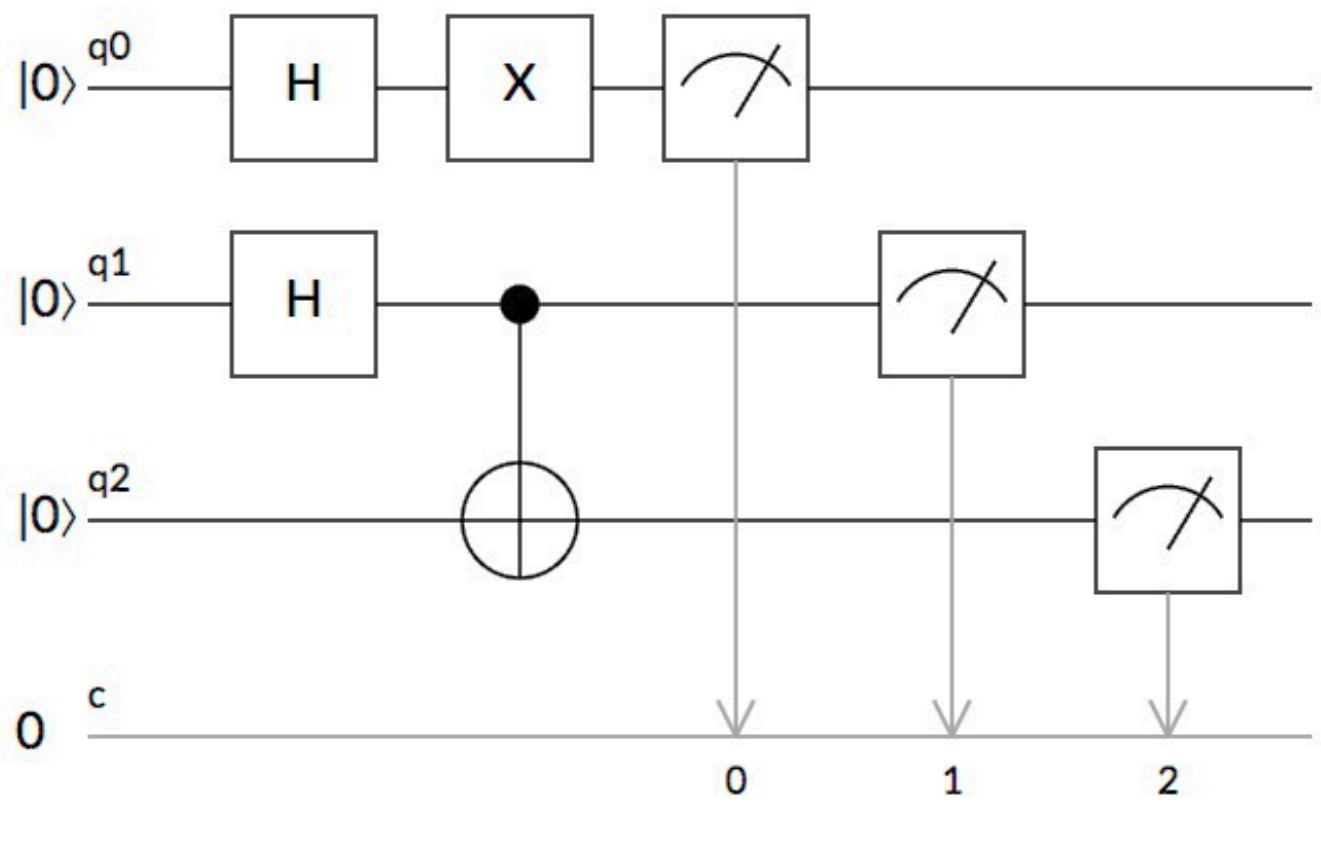

**Figure 12:** A quantum program depicted as a circuit of quantum gates. The squares with dials represent measurements, which are saved on classic registers shown at the bottom line.

Quantum computation gets really interesting with gates that operate on multiple qubits, such as the 'conditional X gate', or 'CX gate'. The CX gate puts two qubits in *entanglement*.

Entanglement establishes a curious correlation between qubits. In practice, the CX gate applies an X gate on a qubit only if the state of another qubit is $|1\rangle$. Thus, the CX gate establishes a dependency of the state of one qubit with the value of another (Figure 13). In practice, any quantum gate can be made conditional and entanglement can take place between more than two qubits.

The Bloch sphere is useful to visualize what happens with a single qubit, but it is not suitable for multiple qubits, in particular when they are entangled. Entangled qubits can no longer be thought of as independent units. They become one quantum entity described by a state vector of its own right on a hypersphere. A hypersphere is an extension of the Bloch sphere to $2^n$ complex dimensions, where $n$ is the number of qubits. Quantum gates perform rotations of a state vector to a new position on this hypersphere. Thus, it is virtually impossible to visualize a system with multiple qubits. There is no better way but to use mathematics to represent quantum systems.

The notation used above to represent quantum states ($|\Psi\rangle, |0\rangle, |1\rangle$), is referred to as *Dirac notation*, which provides an abbreviated way to represent a vector. For instance, $|0\rangle$ and $|1\rangle$ represent the following vectors, respectively:

$$|0\rangle = \begin{bmatrix} 1 \\ 0 \end{bmatrix} \quad \text{and} \quad |1\rangle = \begin{bmatrix} 0 \\ 1 \end{bmatrix}$$

And quantum gates are represented as matrices. For instance, the X gate is represented as:

$$X = \begin{bmatrix} 0 & 1 \\ 1 & 0 \end{bmatrix}$$

Therefore, quantum gate operations are represented as matrix operations; e.g., multiplication of a matrix (gate) by a vector (qubit state). Thus, the application of an X gate to $|0\rangle$ looks like this:

$$X(|0\rangle) = \begin{bmatrix} 0 & 1 \\ 1 & 0 \end{bmatrix} \times \begin{bmatrix} 1 \\ 0 \end{bmatrix} = \begin{bmatrix} 0 \\ 1 \end{bmatrix} = |1\rangle$$

Conversely, the application of an X gate to $|1\rangle$ would therefore is written as follows:

$$X(|1\rangle) = \begin{bmatrix} 0 & 1 \\ 1 & 0 \end{bmatrix} \times \begin{bmatrix} 0 \\ 1 \end{bmatrix} = \begin{bmatrix} 1 \\ 0 \end{bmatrix} = |0\rangle$$

The Hadamard gate has the matrix:

$$H = \begin{bmatrix} \frac{1}{\sqrt{2}} & \frac{1}{\sqrt{2}} \\ \frac{1}{\sqrt{2}} & -\frac{1}{\sqrt{1}} \end{bmatrix} = \frac{1}{\sqrt{2}} \begin{bmatrix} 1 & 1 \\ 1 & -1 \end{bmatrix}$$

As we have seen earlier, the application of the H gate to a qubit pointing to $|0\rangle$ puts it in superposition, right at the equator of the Bloch sphere. This is notated as follows:

$$H(|0\rangle) = \frac{1}{\sqrt{2}}(|0\rangle + |1\rangle)$$

As applied to $|1\rangle$, it also puts it in superposition, but pointing to the opposite direction of the superposition shown above:

$$H(|1\rangle) = \frac{1}{\sqrt{2}}(|0\rangle - |1\rangle)$$

In the preceding equations, the result of $H(|0\rangle)$ and $H(|1\rangle)$ could written as $|+\rangle$ and $|-\rangle$, respectively. In a circuit, we could subsequently apply another gate to $|+\rangle$ or $|-\rangle$, and so on; e.g. $X(|+\rangle)$ = $|+\rangle$.

The Hadamard gate is often used to change the so-called *computational basis* of the qubit. The z-axis $|0\rangle$ and $|1\rangle$ form the *standard basis*. The x-axis $|+\rangle$ and $|-\rangle$ forms the so-called *conjugate basis*. The application of $X(|+\rangle)$ would not have much effect if we measure the qubit in the *standard basis*: it would still probabilistically return 0 or 1. However, it would be different if we were to measure it in the *conjugate basis*; it would deterministically return the value on the opposite side where the vector is aiming to. Another commonly used basis is the *circular basis* (y-axis). A more detailed explanation of different bases and their significance to computation and measurement can be found in [19]. What is important to keep in mind is that changing the basis on which a quantum state is expressed, corresponds to changing the kind of measurement we perform, and so, naturally, it also changes the probabilities of measurement outcomes.

Quantum processing with multiple qubits is represented by means of *tensor vectors*. A tensor vector is the result of the tensor product, represented by the symbol ⊗, of 2 or more vectors. A system of two qubits looks like this $|0\rangle \otimes |0\rangle$, but it is normally abbreviated to $|00\rangle$. However, it is useful to look at the expanded form of the tensor product to follow how it works:

$$|00\rangle = |0\rangle \otimes |0\rangle = \begin{bmatrix} 1 \\ 0 \end{bmatrix} \otimes \begin{bmatrix} 1 \\ 0 \end{bmatrix} = \begin{bmatrix} 1 \times 1 \\ 1 \times 0 \\ 0 \times 1 \\ 0 \times 0 \end{bmatrix} = \begin{bmatrix} 1 \\ 0 \\ 0 \\ 0 \end{bmatrix}$$

Similarly, the other 3 possible states of a 2-qubits system are as follows:

$$|01\rangle = |0\rangle \otimes |1\rangle = \begin{bmatrix} 1 \\ 0 \end{bmatrix} \otimes \begin{bmatrix} 0 \\ 1 \end{bmatrix} = \begin{bmatrix} 1 \times 0 \\ 1 \times 1 \\ 0 \times 0 \\ 0 \times 1 \end{bmatrix} = \begin{bmatrix} 0 \\ 1 \\ 0 \\ 0 \end{bmatrix}$$

$$|10\rangle = |1\rangle \otimes |0\rangle = \begin{bmatrix} 0 \\ 1 \end{bmatrix} \otimes \begin{bmatrix} 1 \\ 0 \end{bmatrix} = \begin{bmatrix} 0 \times 1 \\ 0 \times 0 \\ 1 \times 1 \\ 1 \times 0 \end{bmatrix} = \begin{bmatrix} 0 \\ 0 \\ 1 \\ 0 \end{bmatrix}$$

$$|11\rangle = |1\rangle \otimes |1\rangle = \begin{bmatrix} 0 \\ 1 \end{bmatrix} \otimes \begin{bmatrix} 0 \\ 1 \end{bmatrix} = \begin{bmatrix} 0 \times 0 \\ 0 \times 1 \\ 1 \times 0 \\ 1 \times 1 \end{bmatrix} = \begin{bmatrix} 0 \\ 0 \\ 0 \\ 1 \end{bmatrix}$$

We are now in a position to explain how the CX gate works in more detail. This gate is defined by the matrix:

$$CX = \begin{bmatrix} 1 & 0 & 0 & 0 \\ 0 & 1 & 0 & 0 \\ 0 & 0 & 0 & 1 \\ 0 & 0 & 1 & 0 \end{bmatrix}$$

The application of CX to $|00\rangle$ is represented as:

$$CX(|00\rangle) = \begin{bmatrix} 1 & 0 & 0 & 0 \\ 0 & 1 & 0 & 0 \\ 0 & 0 & 0 & 1 \\ 0 & 0 & 1 & 0 \end{bmatrix} \times \begin{bmatrix} 1 \\ 0 \\ 0 \\ 0 \end{bmatrix} = \begin{bmatrix} 1 \\ 0 \\ 0 \\ 0 \end{bmatrix}$$

The resulting vector is then abbreviated to $|00\rangle$ as show below:

$$\begin{bmatrix} 1 \\ 0 \\ 0 \\ 0 \end{bmatrix} = \begin{bmatrix} 1 \\ 0 \end{bmatrix} \otimes \begin{bmatrix} 1 \\ 0 \end{bmatrix} = |0\rangle \otimes |0\rangle = |00\rangle$$

None that $|00\rangle$ incurred no change because the conditional qubit (the one on the left side of the pair) is not equal to $|1\rangle$. Conversely, should one apply CX to $|10\rangle$, then there is a change to $|11\rangle$, as follows:

$$CX(|10\rangle) = \begin{bmatrix} 1 & 0 & 0 & 0 \\ 0 & 1 & 0 & 0 \\ 0 & 0 & 0 & 1 \\ 0 & 0 & 1 & 0 \end{bmatrix} \times \begin{bmatrix} 0 \\ 0 \\ 1 \\ 0 \end{bmatrix} = \begin{bmatrix} 0 \\ 0 \\ 0 \\ 1 \end{bmatrix}$$

$$\begin{bmatrix} 0 \\ 0 \\ 0 \\ 1 \end{bmatrix} = \begin{bmatrix} 0 \\ 1 \end{bmatrix} \otimes \begin{bmatrix} 0 \\ 1 \end{bmatrix} = |1\rangle \otimes |1\rangle = |11\rangle$$

Table 1 shows the resulting quantum states of CX gate operations, where the first qubit flips only if the second qubit is $1$. Figure 13 illustrates how the CX gate is represented in a circuit diagram. Note that in quantum computing qubit strings are enumerated from the right end of the string to the left: ... $|q_2\rangle \otimes |q_1\rangle \otimes |q_0\rangle$.

| Input | Result |
|---|---|
| $\lvert 00\rangle$ | $\lvert 00\rangle$ |
| $\lvert 01\rangle$ | $\lvert 01\rangle$ |
| $\lvert 10\rangle$ | $\lvert 11\rangle$ |
| $\lvert 11\rangle$ | $\lvert 10\rangle$ |

**Table 1:** CX gate table.

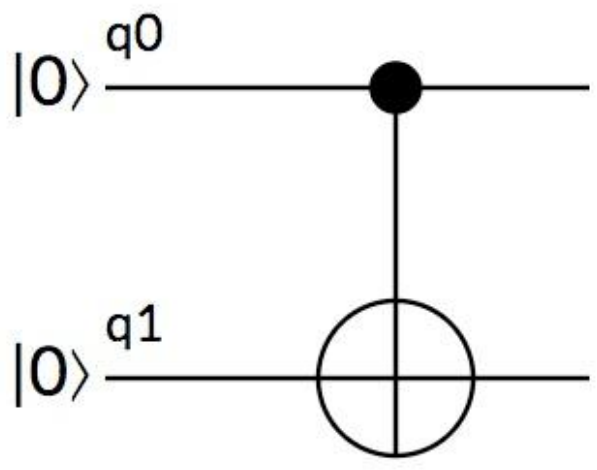

**Figure 13:** The CX gate creates a dependency of the state of one qubit with the state of another. In this case, q1 will be flipped only is q0 is $|1\rangle$.

Another useful conditional gate, which appears on a number of quantum algorithms, is the CCX gate, also known as the *Toffoli gate*, involving three qubits (Figure 14).

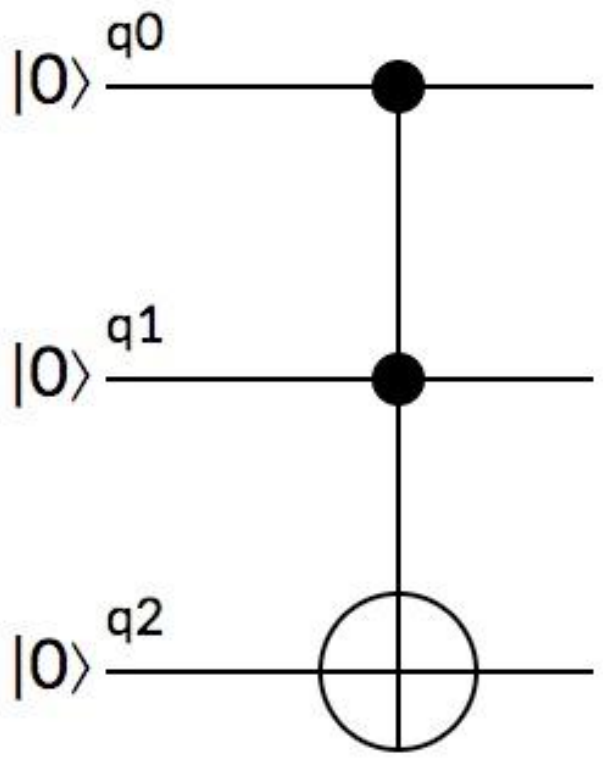

**Figure 14:** The Toffoli gate creates a dependency of the state of one qubit with the state of two others.

Table 2 shows resulting quantum states of the Toffoli gate: the first qubit flips only if the second and the third are 1.

| **Input** | **Result** |
|---|---|
| $\|000\rangle$ | $\|000\rangle$ |
| $\|001\rangle$ | $\|001\rangle$ |
| $\|010\rangle$ | $\|010\rangle$ |
| $\|011\rangle$ | $\|011\rangle$ |
| $\|100\rangle$ | $\|100\rangle$ |
| $\|101\rangle$ | $\|101\rangle$ |
| $\|110\rangle$ | $\|111\rangle$ |
| $\|111\rangle$ | $\|110\rangle$ |

**Table 2:** Toffoli gate table.

The equation for describing a 2-qubits system $|q_1\rangle \otimes |q_0\rangle$ combines two state vectors $|\Psi\rangle$ and $|\Phi\rangle$ as follows. Consider:

$$|\Psi\rangle = \alpha_1|0\rangle + \alpha_2|1\rangle \text{ for } q_0$$

$$|\Phi\rangle = \beta_1|0\rangle + \beta_2|1\rangle \text{ for } q_1$$

Then,

$$|\Psi\rangle \otimes |\Phi\rangle = \alpha_0\beta_0|00\rangle + \alpha_0\beta_1|01\rangle + \alpha_1\beta_0|10\rangle + \alpha_1\beta_1|11\rangle$$

The above represents a new quantum state with four amplitude coefficients, which can be written as:

$$|A\rangle = \alpha_0|00\rangle + \alpha_1|01\rangle + \alpha_2|10\rangle + \alpha_3|11\rangle$$

Consider this equation:

$$|\Psi\rangle = \tfrac{1}{2}|00\rangle + \tfrac{1}{2}|01\rangle + \tfrac{1}{2}|10\rangle + \tfrac{1}{2}|11\rangle$$

The above is saying that each of the four quantum states have equal probability of 25% each of being returned.

Now, it should be straightforward to work out how to describe quantum systems with more qubits. For instance, a system with four qubits looks like this:

$$\begin{aligned}|B\rangle = \ &\beta_0|0000\rangle + \beta_1|0001\rangle + \beta_2|0010\rangle + \beta_3|0011\rangle + \\ &\beta_4|0100\rangle + \beta_5|0101\rangle + \beta_6|0110\rangle + \beta_7|0111\rangle + \\ &\beta_8|1000\rangle + \beta_9|1001\rangle + \beta_{10}|1010\rangle + \beta_{11}|1011\rangle + \\ &\beta_{12}|1100\rangle + \beta_{13}|1101\rangle + \beta_{14}|1110\rangle + \beta_{15}|1111\rangle\end{aligned}$$

A linear increase of the number of qubits extends the capacity of representing information on a quantum computer exponentially. With qubits in superposition, a quantum computer can handle all possible values of some input data simultaneously. This endows the machine with massive parallelism. However, we do not have access to the information until the qubits are measured.

Quantum algorithms require a different way of thinking than the way one normally approaches programming; for instance, it is not possible to store quantum states on a working memory for accessing later in the algorithm. This is due to the so-called *non-cloning principle* of quantum physics: it is impossible to make a copy of a quantum system. It is possible, however, to move the state of a set of qubits to another set of qubits, but in effect this deletes the information from the original qubits. To program a quantum computer requires manipulations of qubits so that the states that correspond to the desired outcome have a much higher probability of being measured than all the other possibilities.

Decoherence is problematic because it poses limitations on the number of successive gates that can be used in a circuit; that is, the *circuit depth*. The more gates we use, the higher the chances that the qubits will decohere. And this inevitably causes errors. In particular, running a circuit which is deeper than the critical depth for which a quantum device can maintain coherence will result in measurement outcomes sampled from an effectively classical distribution, sadly defeating the whole purpose of using a quantum computer in the first place. At the time of writing, QPUs do not have more than a few dozen qubits and are unable to maintain a desired quantum state for longer than a few milliseconds.

One way to mitigate errors is to run the algorithms (which should not be too deep) many times and then select the result that appeared most. Additional post processing on the measurement outcomes that tries to undo the effect of the noise by solving an inverse problem can also be carried out. Increasingly sophisticated error correction methods are also being developed. And better hardware technology is also developing fast. But as stated above, highly fault-tolerant quantum computation is still a long way from being realised.

## 5 Quantum Vocal Synthesizer

This section introduces an interactive vocal synthesizer with parameters determined by a *quantum hyper-die*. The system listens to a tune chanted on a microphone and counts the number of notes it can detect in the signal. Then, it synthesizes the same amount of sounds as the number of notes that it counted in the tune. The synthesized vocal sounds are not intended to imitate the listened tune. Rather, they are 'quantum' responses, whose make-up is defined by the quantum hyper-die.

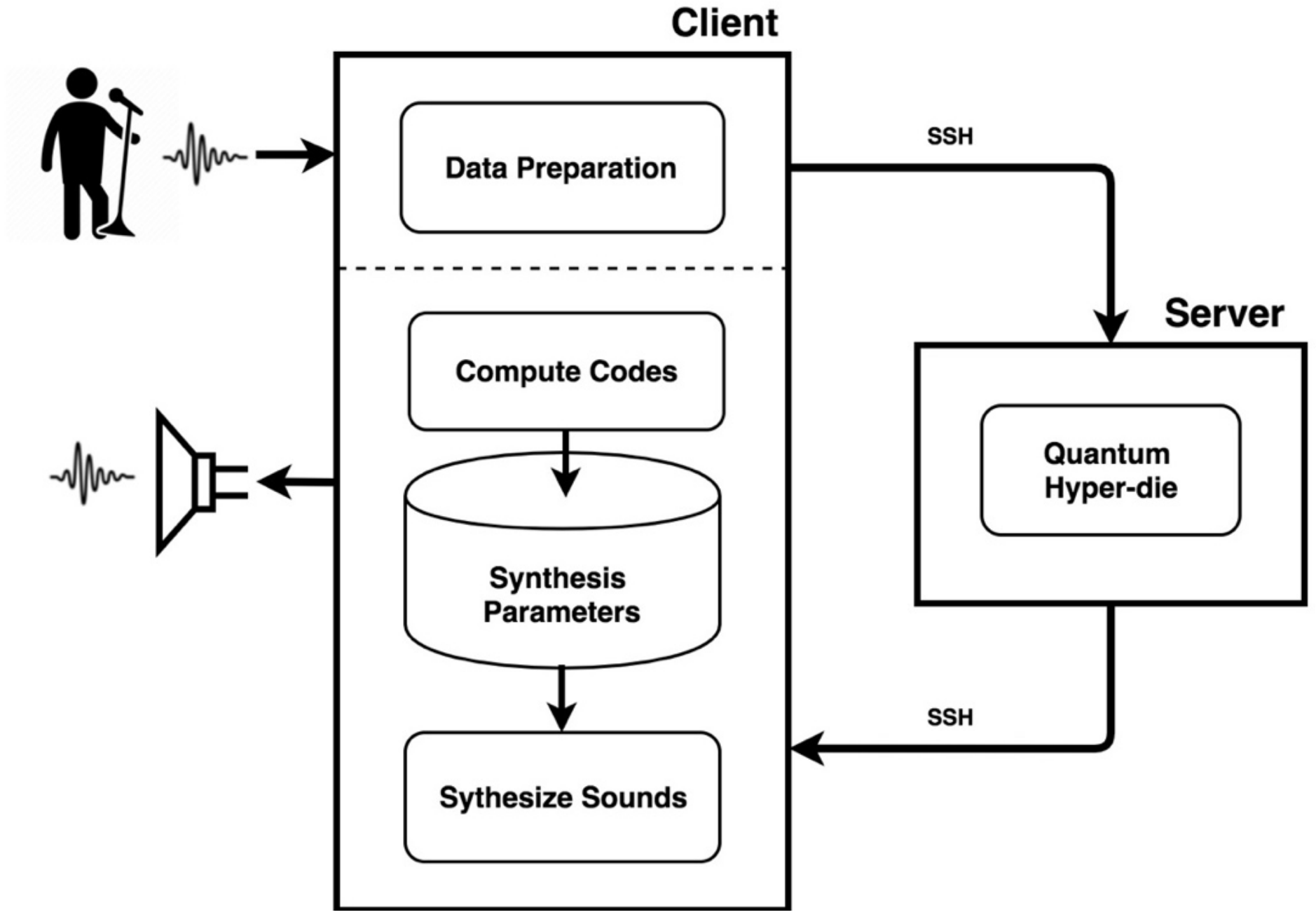

**Figure 15:** The interactive quantum vocal system architecture.

The system comprises two components connected through the Internet using the SSH (Secure Shell) protocol: a client and server (Figure 15). The client operates on a standard laptop computer and the server on a Rigetti's Forest quantum computer, located in Berkeley, California.

The server runs the hyper-die quantum circuit depicted in Figure 19 and sends measurements to the client. The client takes care of analysing the chanted tune, preparing data to set up the server, and synthesizing sounds based on the results of the measurements.

The system is programmed in Python and uses pyQuil, a Python library developed by Rigetti to write quantum programs [4]. The core of the vocal synthesiser is implemented using the programming language Csound [10]. The Csound code is called from within Python.

The audio spectrum of singing human voice has the appearance of a pattern of 'hills and valleys'. The 'hills' are referred to as *formants* (Figure 16). A vocal sound usually has between three to five distinct formants. Each of them comprises a set of sound partials.

A formant is described by a frequency, which is the frequency of its most prominent partial, and an amplitude, which is the energy of this frequency. A third descriptor is the formant's bandwidth, which is the width of the 'hill'. It is calculated as the difference between the highest and the lowest frequencies in the formant set. Frequencies and bandwidths are quantified in Hz and amplitudes in dB.

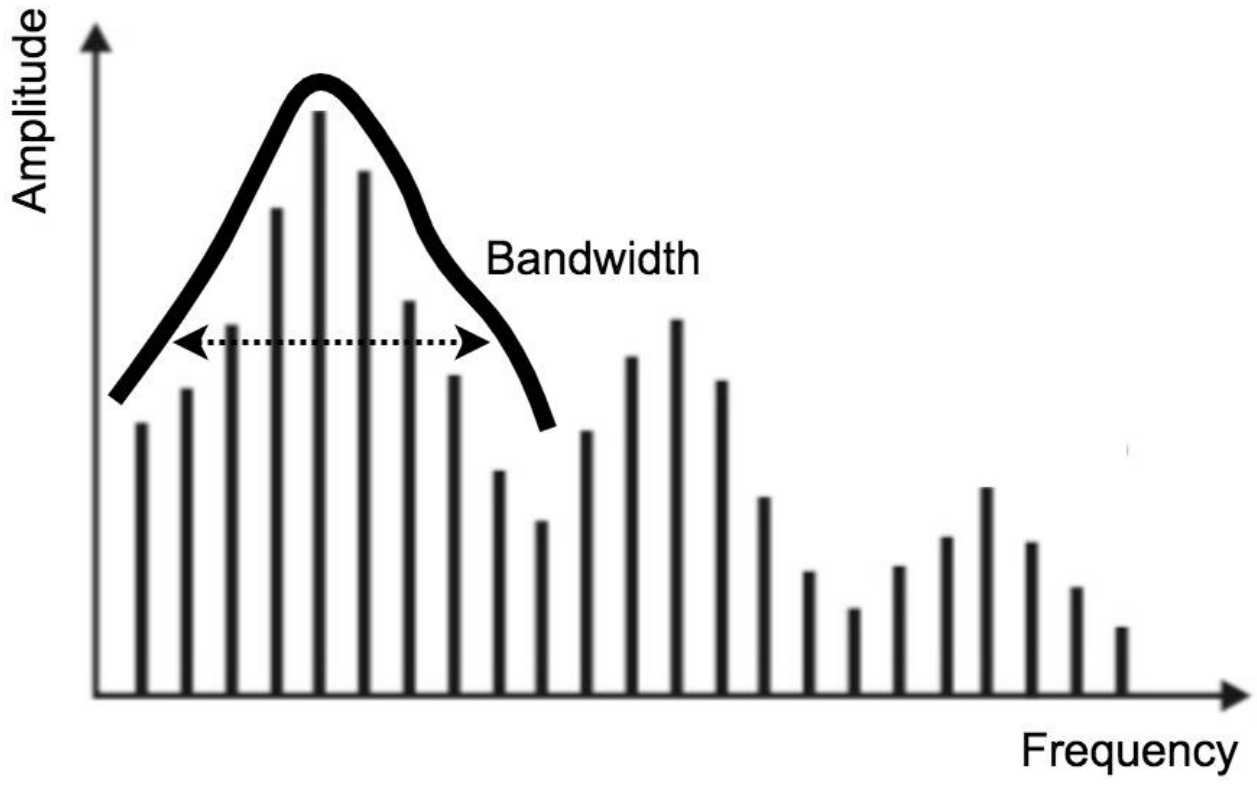

**Figure 16:** Generalized spectrum of the human voice with three formants.

Normally, the first three formants of a vocal sound characterise its phonetic timbre. For instance, they define whether a vowel is open (e.g., as in /a/ in the word 'back') or close (e.g., as in /o/ in the word 'too'); e.g., the frequency of the first formant of an open vowel is higher than that of a close vowel.

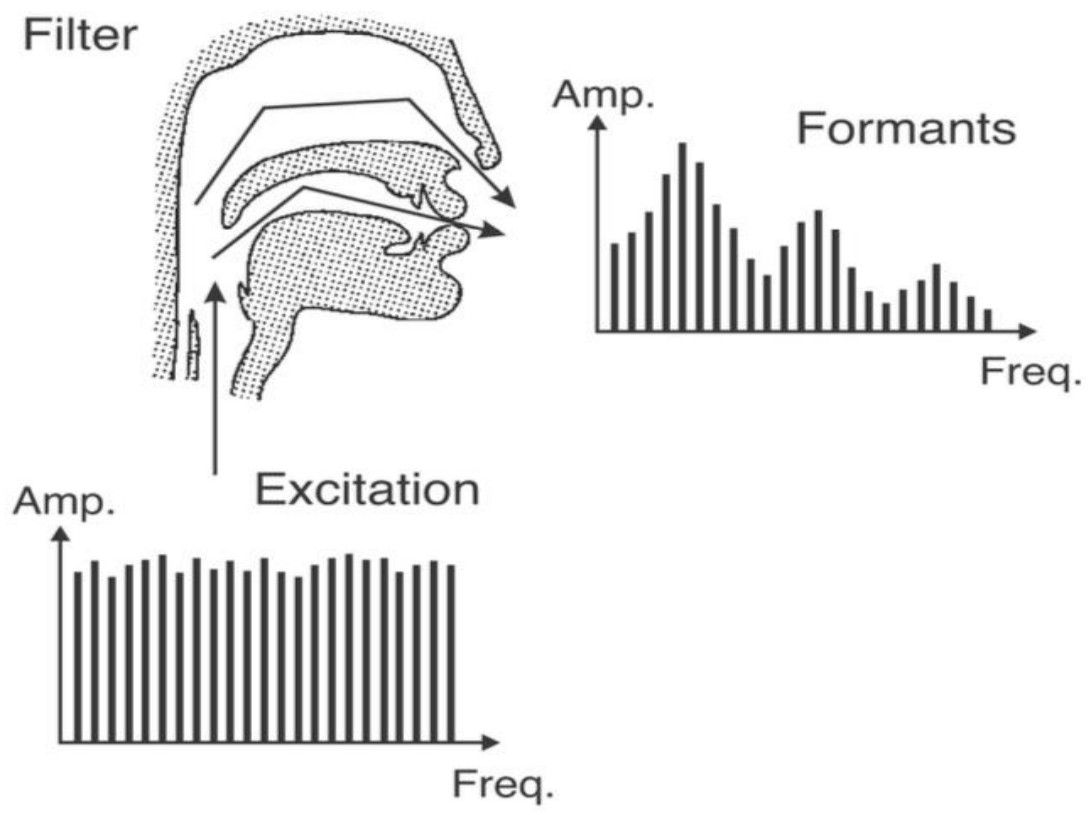

**Figure 17:** Vocal system.

Traditionally, the vocal system has been modelled as a system consisting of two modules: a source module and a resonator module. The source module produces an excitation signal. Then, this signal is altered by the acoustic response of the resonator. The excitation signal is intended to simulate the waveform produced by the vibration of the glottis. The throat, nasal cavity and mouth function as resonating chambers whereby particular frequencies of the excitation signal are emphasised and others are attenuated (Figure 17).

There are a number of methods to synthesise simulations of singing voice [20]. The synthesis method used here is known as FOF, which is an acronym for *Fonctions d'Onde Formantique*, or Formant Wave Functions, in English [21].

The core of the synthesizer comprises five formant generators in parallel to produce five formants (Figure 18). Each FOF generator requires 15 input parameters to produce a formant. A detailed explanation of FOF is beyond the scope of this chapter. For the sake of simplicity, we shall focus here on three parameters only: formant's frequency (`fq`), formant's amplitude (`amp`) and formant's bandwidth (`bw`).

Each formant generator is controlled by three linear functions. The functions vary the generator's input frequency, amplitude and bandwidth, from initial to end values. For instance, `fq1s` is the starting frequency value for generator Formant 1, whereas `fq1e` is the ending value. These variations are continuous and last through the entire duration of the sound.

The outputs from the oscillators are summed and a vibrato generator is applied, which is also controlled by linear functions. Vibrato renders the result more realistic to our ears. Then, an ADSR (short for attack, decay, sustain and release) function shapes the overall amplitude of the sound. Other parameters for the synthesizer are the resulting sound's fundamental frequency, or pitch (`fnd`), its loudness (`ldns`) and its duration (`dur`).

The quantum hyper-die is a simple quantum circuit that puts 9 qubits in superposition and measures them (Figure 19). This results in a set of 9 measurements, which are processed by the client to produce codes of three bits each. These codes are used to retrieve synthesis parameter values from a database. The database contains valid values for all synthesis parameters shown in Figure 18, plus other ones that are not shown.

For instance, consider the list of measurements [$c_8$, $c_7$, $c_6$, $c_5$, $c_4$, $c_3$, $c_2$, $c_1$, $c_0$]. Codes are produced by combining three elements from the measurements list according to a bespoke combinatorial formula. For example, ($c_8$ $c_7$ $c_6$), ($c_6$ $c_7$ $c_8$), ($c_5$ $c_4$ $c_3$), ($c_3$ $c_4$ $c_5$), ($c_2$ $c_1$ $c_0$), ($c_0$ $c_1$ $c_2$) and so forth. Given a list of synthesis parameter values [$p_0$, $p_1$, $p_2$, $p_3$, $p_4$, $p_5$, $p_6$, $p_7$], the decimal value of a code gives an index to retrieve a value for the synthesizer. For instance, the code (0 1 0), which yields the decimal number 2, would retrieve $p_2$.

Each synthesis parameter is coupled with a unique code formation for retrieval. For instance, ($c_8$ $c_7$ $c_6$) is coupled with the starting frequency for the first formant (`fq1s`) and ($c_6$ $c_7$ $c_8$) with the ending frequency for the first formant (`fq1e`). And ($c_5$ $c_4$ $c_3$) is coupled with the starting frequency for the second formant (`fq2s`), and so on. The database holds different lists of synthesis parameters, which can be customised.

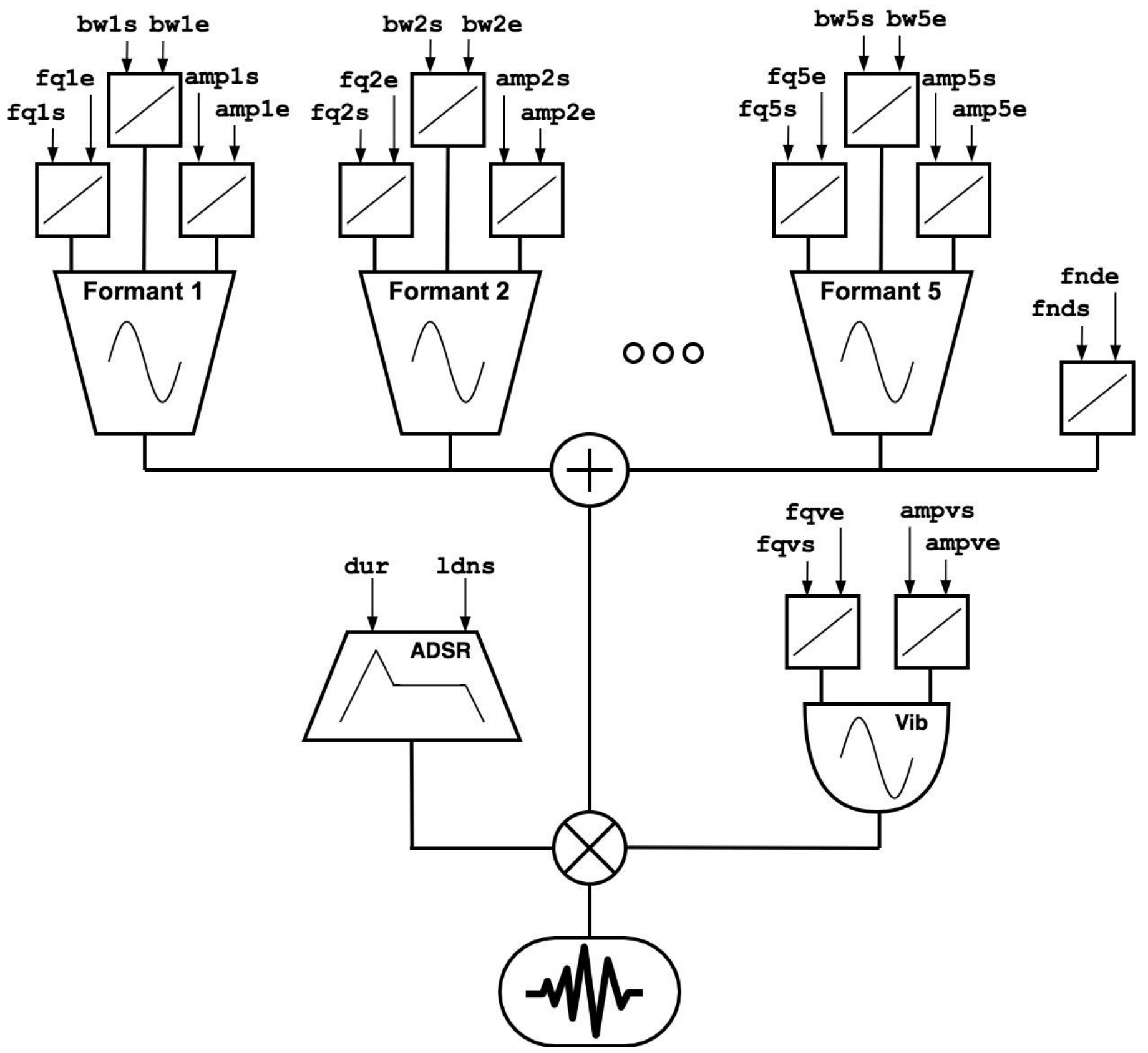

**Figure 18:** Synthesizer's layout.

As an illustration, let us consider a simple database with the following lists of fundamental frequencies and durations, and frequencies, amplitudes and bandwidths for formants 1, 2 and 3:

fnd = [277.2, 185.0, 207.6, 415.3, 155.6, 311.2, 369.9, 233.1]
dur = [3.25, 2.0, 2.75, 4.0, 1.5, 3.75, 2.5, 4.5]
fq1 = [310.0, 270.0, 290.0, 350.0, 650.0, 400.0, 430.0, 470.0]
fq2 = [600.0, 1150.0, 800.0, 1870.0, 1080.0, 1620.0, 1700.0, 1040.0]
fq3 = [2250.0, 2100.0, 2800.0, 2650.0, 2500.0, 2900.0, 2600.0, 2750.0]
amp1 = [0.0, 0.0, 0.0, 0.0, 0.0, 0.0, 0.0, 0.0]
amp2 = [-15, -7, -11, -6, -14, -9, -20, -30]
amp3 = [-9, -21, -12, -32, -17, -16, -10, -18]
bw1 = [35, 60, 45, 70, 80, 75, 58, 85]
bw2 = [65, 70, 90, 75, 83, 95, 60, 87]
bw3 = [128, 115, 110, 112, 98, 104, 124, 120]

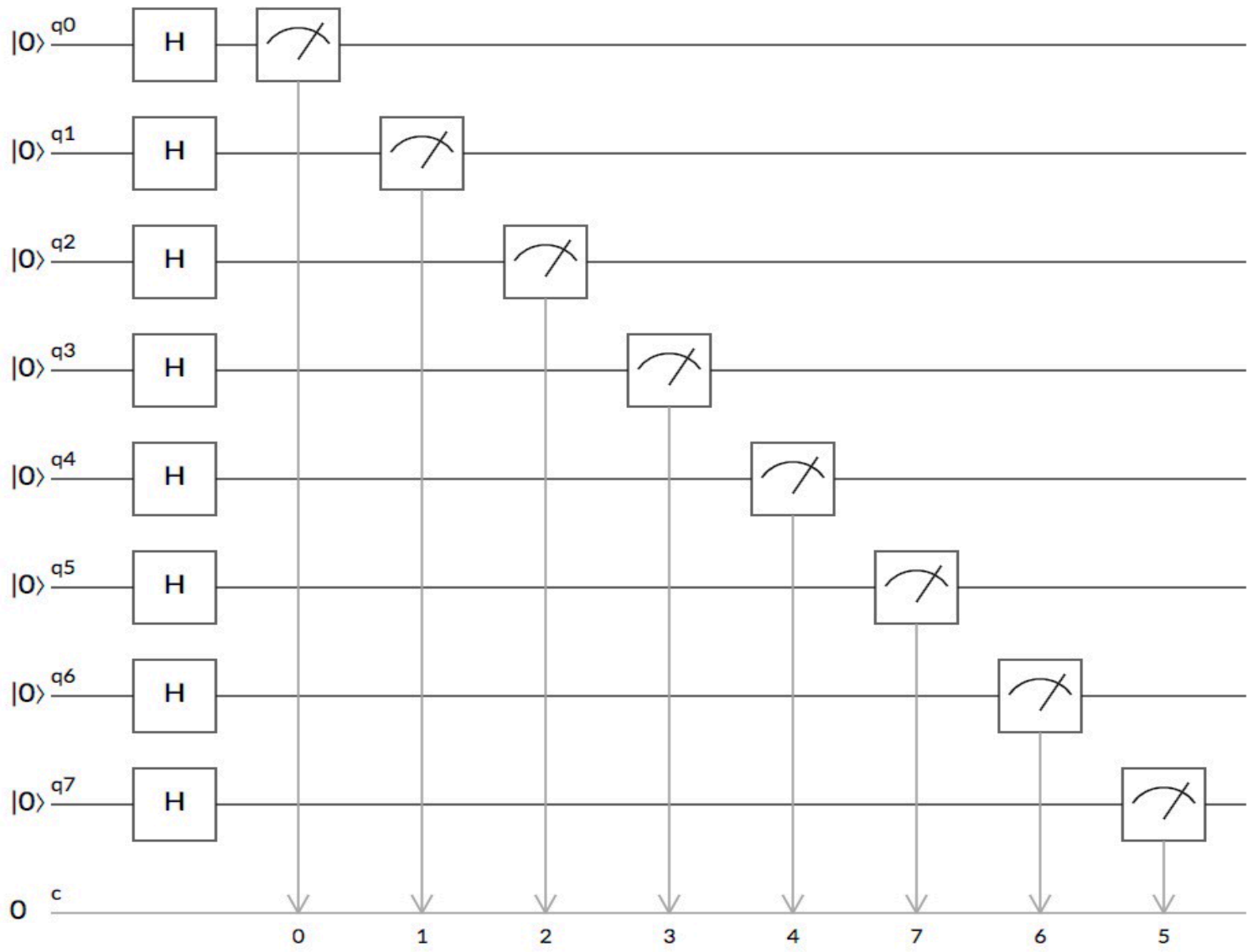

**Figure 19:** The quantum hyper-die circuit. Hadamard gates put all 9 qubits in superposition.

| Code | Binary | Decimal | Parameter | Retrieved Value |
|---|---|---|---|---|
| ($c_8$ $c_7$ $c_6$) | 000 | 0 | `fq1s` | 310.0 Hz |
| ($c_6$ $c_7$ $c_8$) | 000 | 0 | `fq1e` | 310.0 Hz |
| ($c_5$ $c_4$ $c_3$) | 001 | 1 | `fq2s` | 1150.0 Hz |
| ($c_3$ $c_4$ $c_5$) | 100 | 4 | `fq2e` | 1080.0 Hz |
| ($c_2$ $c_1$ $c_0$) | 001 | 1 | `fq3s` | 2100.0 Hz |
| ($c_0$ $c_1$ $c_2$) | 100 | 4 | `fq3e` | 2500.0 Hz |
| ($c_7$ $c_6$ $c_5$) | 000 | 0 | `amp1s` | 0.0 dB |
| ($c_5$ $c_6$ $c_7$) | 000 | 0 | `amp1e` | 0.0 dB |
| ($c_4$ $c_3$ $c_2$) | 010 | 2 | `amp2s` | -11 dB |
| ($c_2$ $c_3$ $c_4$) | 010 | 2 | `amp2e` | -11 dB |
| ($c_8$ $c_5$ $c_2$) | 000 | 0 | `amp3s` | -9 dB |
| ($c_2$ $c_5$ $c_8$) | 000 | 0 | `amp3e` | -9 dB |
| ($c_7$ $c_4$ $c_3$) | 001 | 1 | `bw1s` | 60 Hz |
| ($c_3$ $c_4$ $c_7$) | 100 | 4 | `bw1e` | 80 Hz |
| ($c_6$ $c_3$ $c_0$) | 011 | 3 | `bw2s` | 75 Hz |
| ($c_0$ $c_3$ $c_6$) | 110 | 6 | `bw2e` | 60 Hz |
| ($c_8$ $c_7$ $c_0$) | 001 | 1 | `bw3s` | 115 Hz |
| ($c_0$ $c_7$ $c_8$) | 100 | 4 | `bw3e` | 98 Hz |
| ($c_8$ $c_1$ $c_0$) | 001 | 1 | `fnds` | 185.0 Hz |
| ($c_0$ $c_1$ $c_8$) | 100 | 4 | `fnde` | 155.6 Hz |
| ($c_5$ $c_3$ $c_1$) | 010 | 2 | `dur` | 2.75 secs |

**Table 3:** Retrieved synthesis parameters with codes produced from quantum measurements.

For this example, the server returned the following measurements: [0, 0, 0, 0, 0, 1, 0, 0, 1].

Then, the system produces the respective codes and retrieves the parameters for the synthesiser. For instance, the code ($x_0$ $x_1$ $x_2$) is equal to 000. Therefore, it retrieves the first element of fq1 for the starting frequency of the first formant generator (`fg1s`), which is 310.0 Hz. Table 3 shows the retrieved values for the aforementioned parameters.

Figure 20 shows a spectrogram snapshot taken at two seconds in the sound, showing three prominent formants, whose frequencies match the input parameters yielded by the quantum hyper-die.

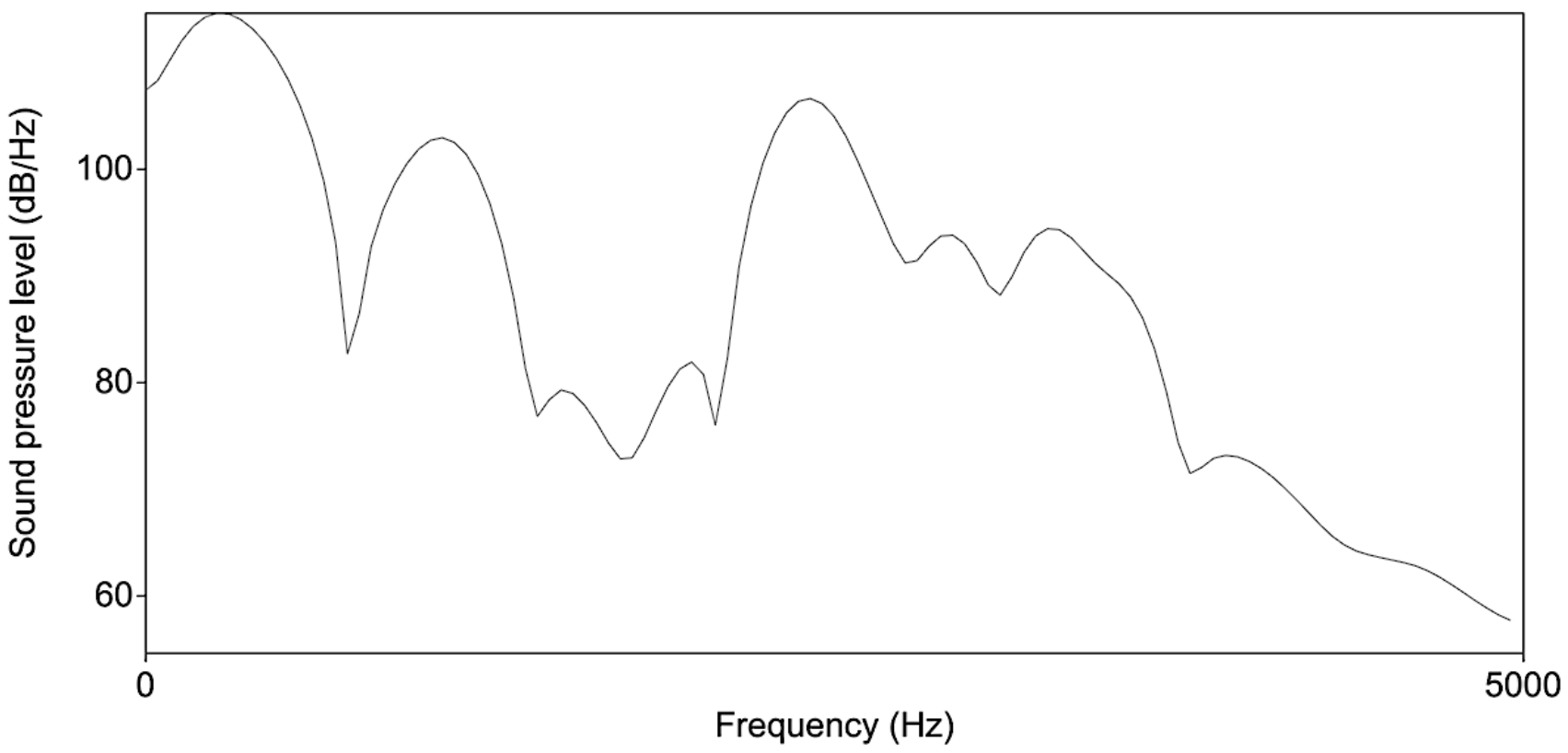

**Figure 20:** Spectrogram analysis snapshot taken at two seconds.

Note that the synthesizer reads amplitudes in terms of attenuations from a hard-coded reference dB value. As a convention, the first formant is always set to this reference, hence the values in the amp1 list are all 0.0 dB. The other amplitudes are calculated by subtracting the negative values from this reference.

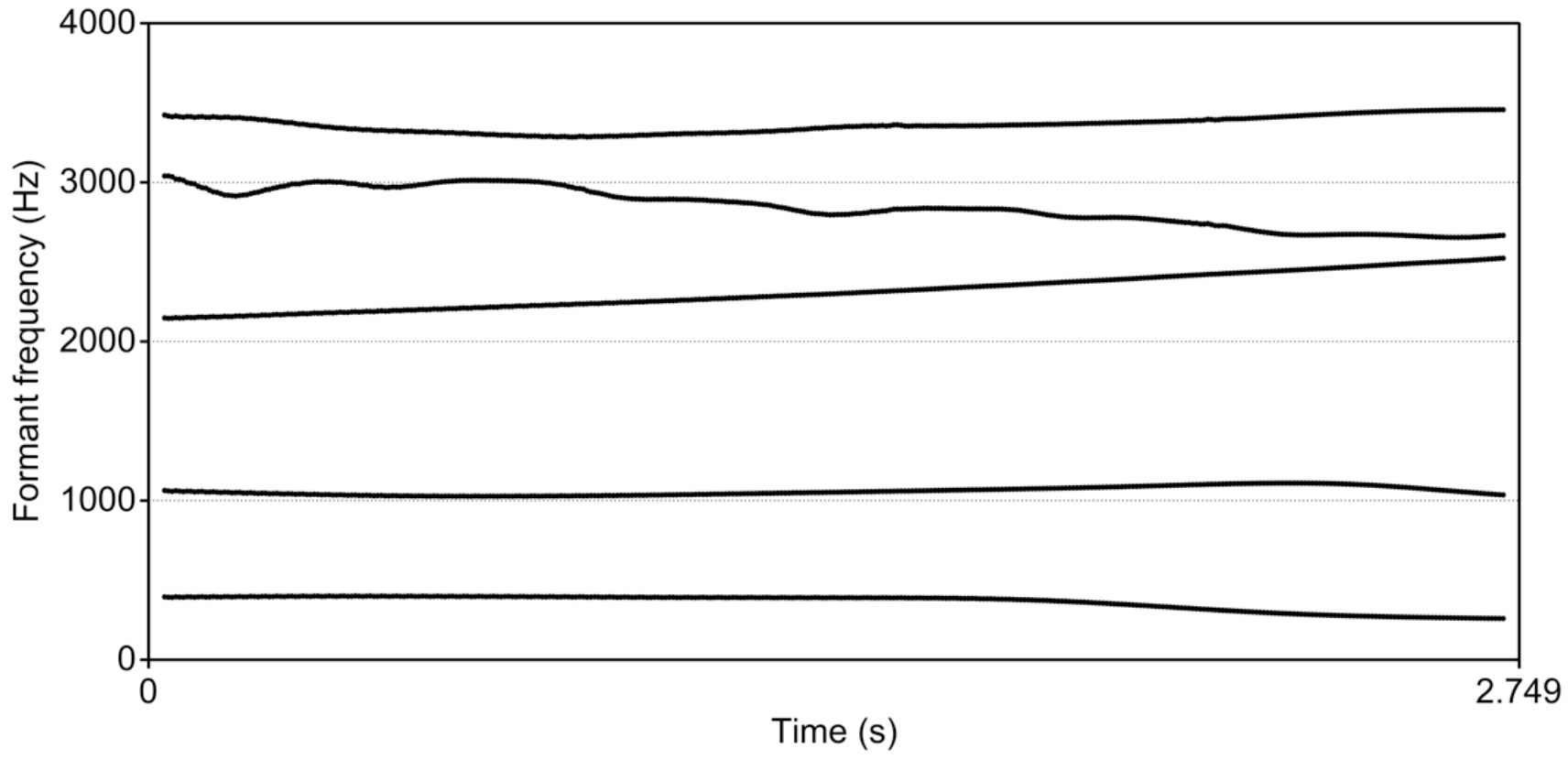

**Figure 21:** Formant analysis revealing five salient spectral components in the synthesised sound.

Figure 21 shows a formant analysis graph of the synthesized sound. Each line corresponds to a formant detected in the signal. Note the first formant is practically constant at 310 Hz throughout the duration of the sound. The third formant, however, raised slightly; i.e., from 1080.0 Hz to 2100.0 Hz.

## 6 Quantum Walk Sequencer

This section presents a system that generates sequences of musical notes using a quantum walk algorithm. For a detailed discussion on quantum random walks, please refer to [22].

As with the quantum vocal synthesiser above, the system consists of two components: a client and a server (Figure 22). The server runs the quantum random walk circuit and sends a list of measurements to the client. Then, the client translates those measurements into a sequence of musical notes, which are encoded as MIDI information [23]. MIDI is a protocol that allows computers, musical instruments and other hardware to communicate. The difference of encoding the results with MIDI rather than synthesizing sounds is that we can connect third party music software to visualize or play back the music.

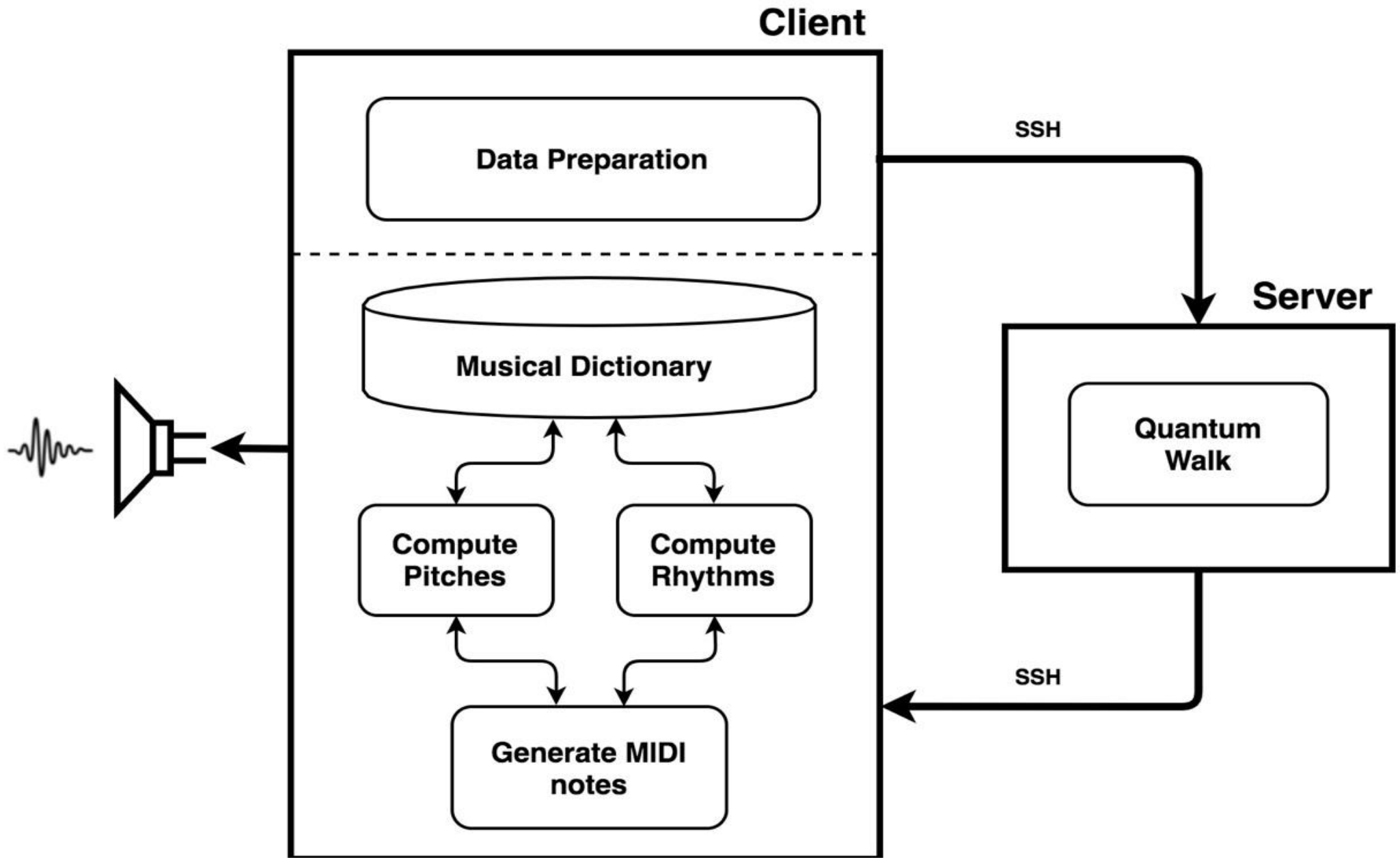

**Figure 22:** The quantum walk music system architecture.

As we have briefly seen earlier in this chapter, in a random walk algorithm, a "walker" starts on a certain node of a graph and has an equal probability of travelling through any connected edge to an adjacent node. This process is then repeated as many times as required. The nodes can represent tasks to be performed once the walker lands on them, or information to handle; e.g., a musical note to be played or appended to a list.

In classical random walk, the walker inhabits a definite node at any one moment in time. But in quantum walk, it will be in a superposition of all nodes it can possibly visit in a given moment. Metaphorically, we could say that the walker is on all viable nodes simultaneously, until we observe it.

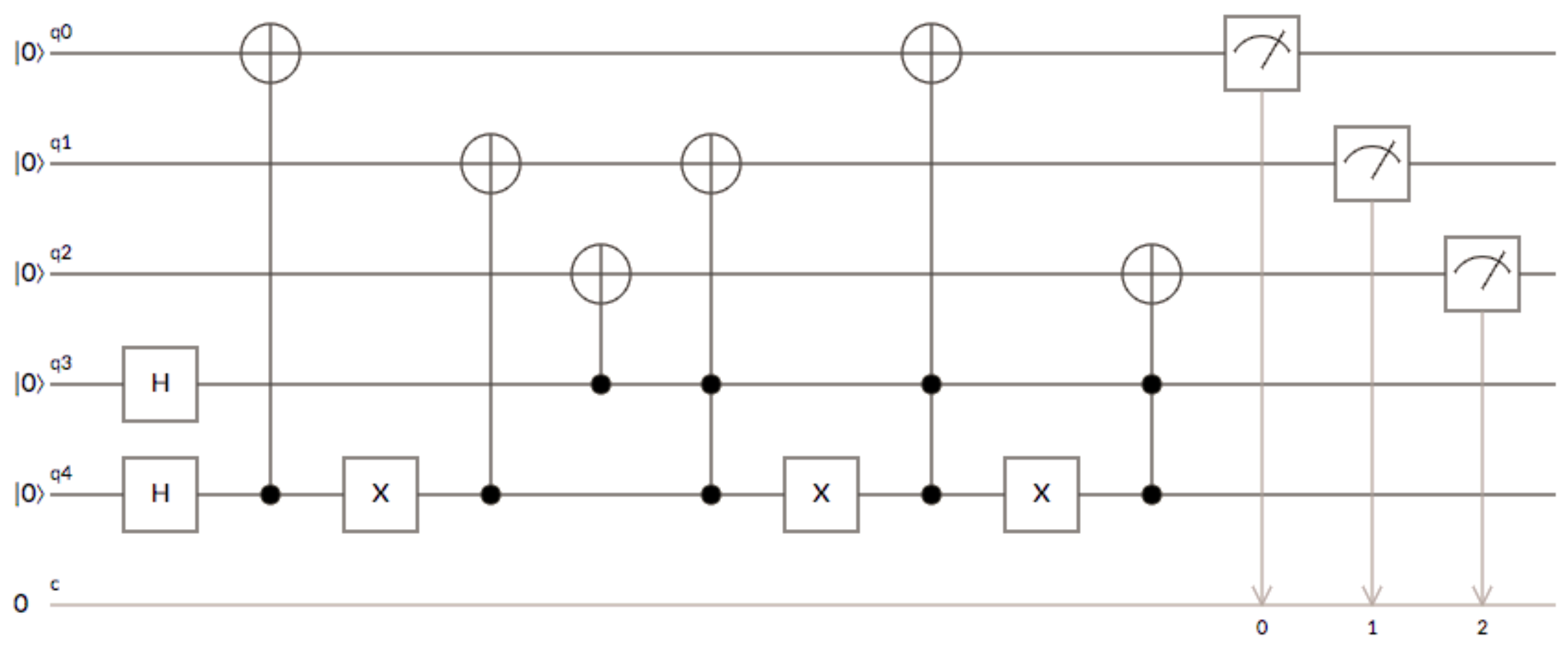

**Figure 23:** Quantum walk circuit.

The circuit (Figure 23) was designed to walk through the edges of a cube to visit eight vertices, each of which is represented as a three bits long binary number (Figure 24). The circuit uses five qubits: three (q0, q1, and q2) to encode the eight vertices of the cube {000, 001, …, 111} and two (q3 and q4) to encode the possible routes that the walker can take from a given vertex, one of which is to stay put. The diagram shows a sequence of operations, the first of which are two H gates applied to q3 and q4, followed by a CX with q4 and q0, and so on.

We refer to the first three qubits as *input qubits* and the last two as *die qubits*. The die qubits act as controls for X gates to invert the state of input qubits.

For every vertex on the cube, the edges connect three neighbouring vertices whose codes differ by changing only one bit of the origin's code. For instance, vertex 111 is connected to 110, 101 and 011. Therefore, upon measurement the system returns one of four possible outputs:

- the original input with inverted q0
- the original input with inverted q1
- the original input with inverted q2
- the original input unchanged

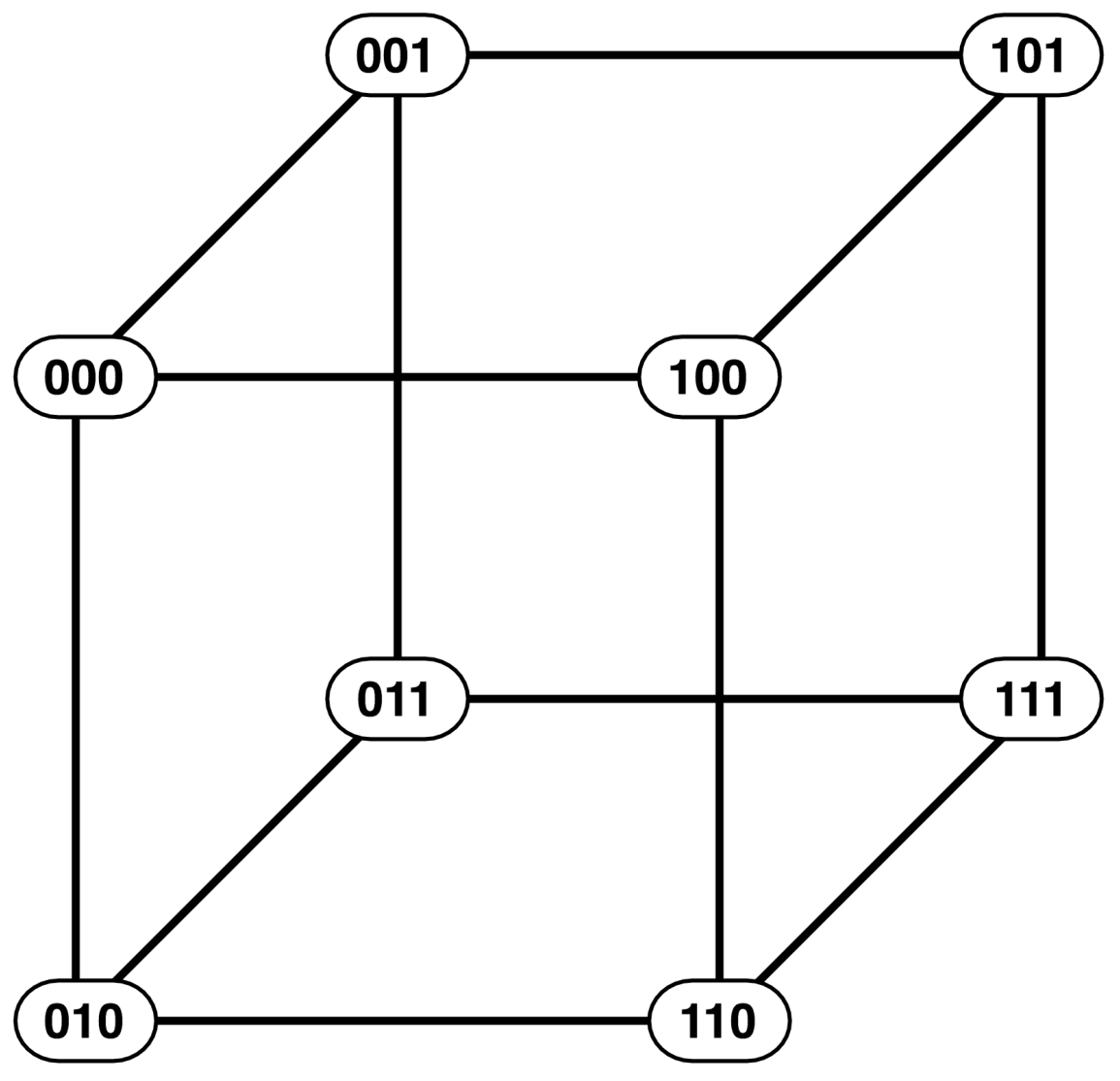

**Figure 24:** Cube representation of the quantum walk routes and nodes.

The quantum walk algorithm runs as follows: the input qubits are armed with the state representing a node of departure and two die qubits are armed in superposition (H gate). Then, the input qubits are measured and the results are stored in a classical memory. This causes the whole system to decohere. Depending on the values yielded by the die qubits, the conditional gates will invert the input qubits accordingly. Note that we measure and store only input qubits; the value of the die can be lost. The result of the measurement is then used to arm the input qubits for the next step of the walk, and the cycle continues for a number of steps. The number of steps is established at the initial data preparation stage.

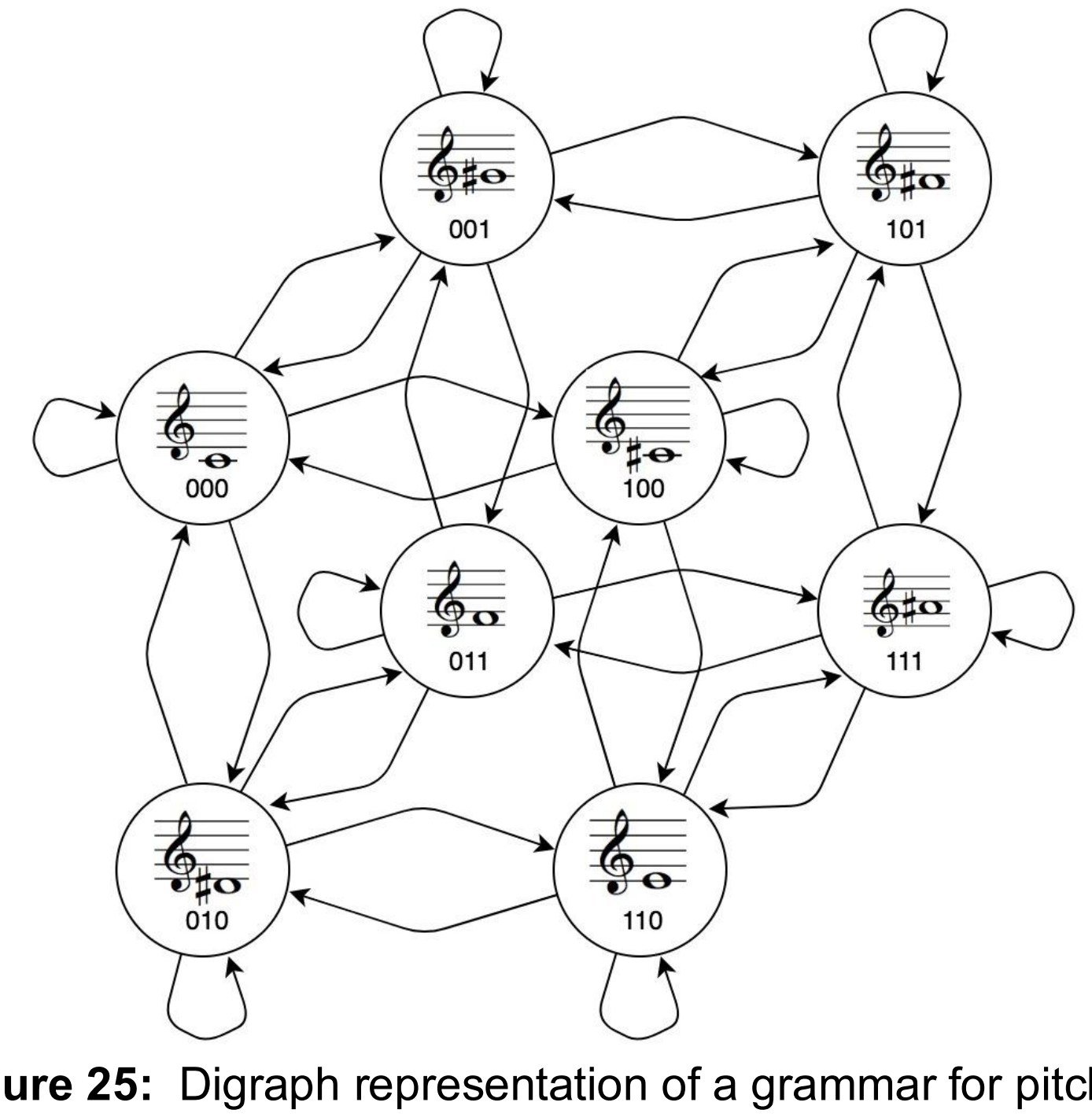

**Figure 25:** Digraph representation of a grammar for pitches.

As a trace table example, let us assume the following input: 001, where q0 is armed to $|0\rangle$, q1 to $|0\rangle$ and q2 to $|1\rangle$. Upon measurement, let us suppose that the die yielded q3 = 0 and q4 = 1. The second operation on the circuit diagram is a CX gate where q4 acts a conditional to invert q0. Right at the second operation the state vector of q0 is inverted because q4 = 1. As the rest of the circuit does not incur any further action on the input qubits, the system returns 101. Should the dice have yielded q3 = 0 and q4 = 0 instead, then the fourth operation would have inverted q1. The X gate (third operation) would have inverted q4, which would subsequently act as a conditional to invert q1. The result in this case would have been 011.

The cube in Figure 24 functions as an underlying common abstract representation of simple musical grammars, whose isomorphic digraphs are shown in Figures 25 and 26. One of the grammars encodes rules for sequencing pitches (Figure 25) and the other rules for sequencing durations of musical notes (Figure 26), respectively. The system holds musical dictionaries associating vertices with pitches and note durations.

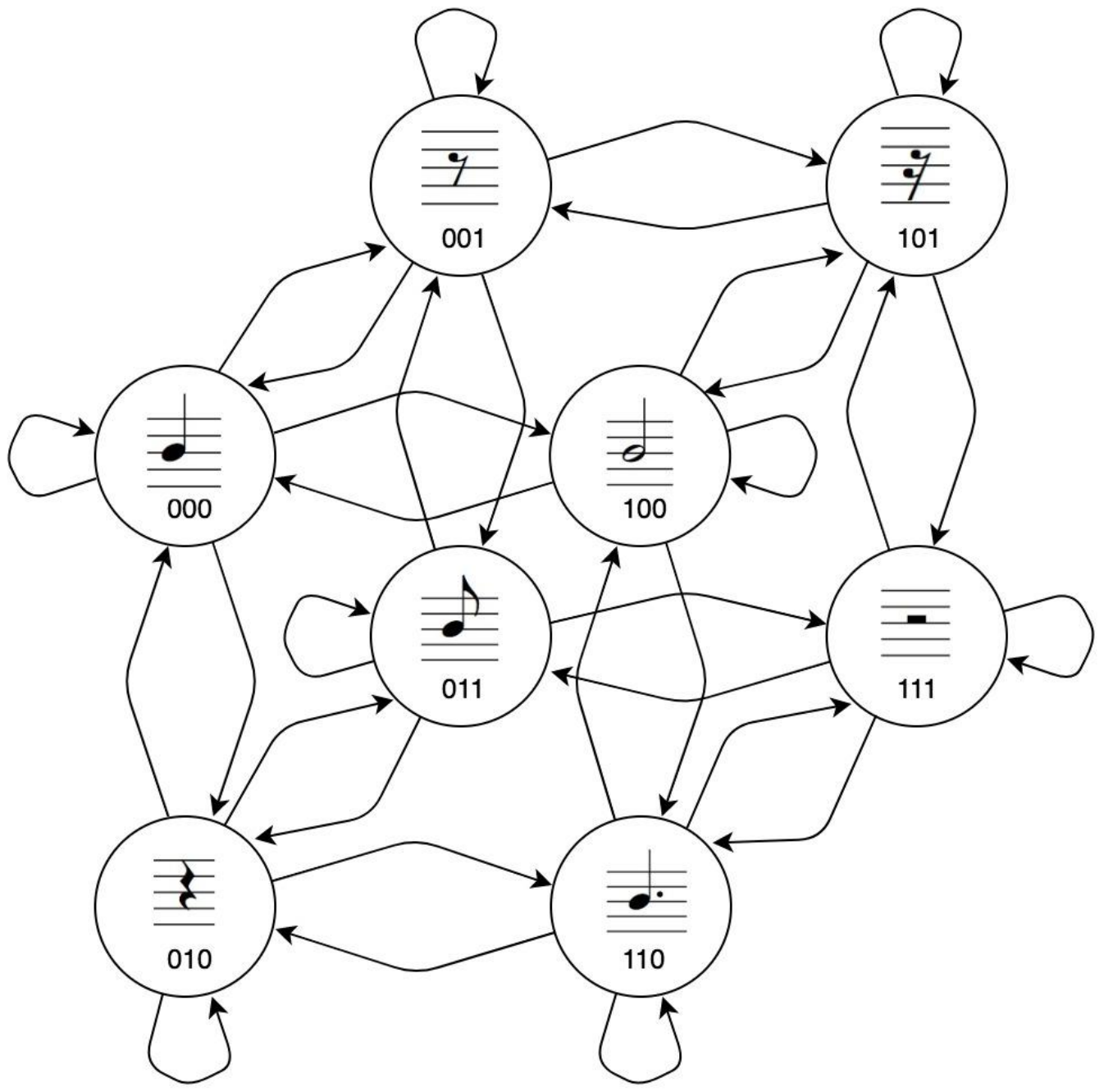

**Figure 26:** Digraph representation of a grammar of durations.

In order to generate a music sequence, the system starts with a given note; for instance, a half note C4, whose codes for pitch and duration are 000 and 100, respectively. This initial note is given to the system at the data preparation stage. Then, for every new note the server runs the quantum walk circuit twice, once with input qubits armed with the code for pitch and then armed with the code for duration. The results from the measurements are then used to establish the next note. For instance, if the first run goes to 001 and the second to 000, then the resulting note is quarter note G4 sharp. The measurements are used to re-arm the circuit for the next note and so on.

An important component of music is silence. Note in Figure 26 that the grammar for durations includes 4 pauses: 001, 010, 101 and 111. When the walker lands on a pause, the pitch of the note is discarded and a silence takes place for the duration of the respective pause. The dictionaries of notes and durations are modifiable and there are tools to change the dictionary during the generation of a sequence. The length of the sequence and the dictionaries are set up at the data preparation stage.

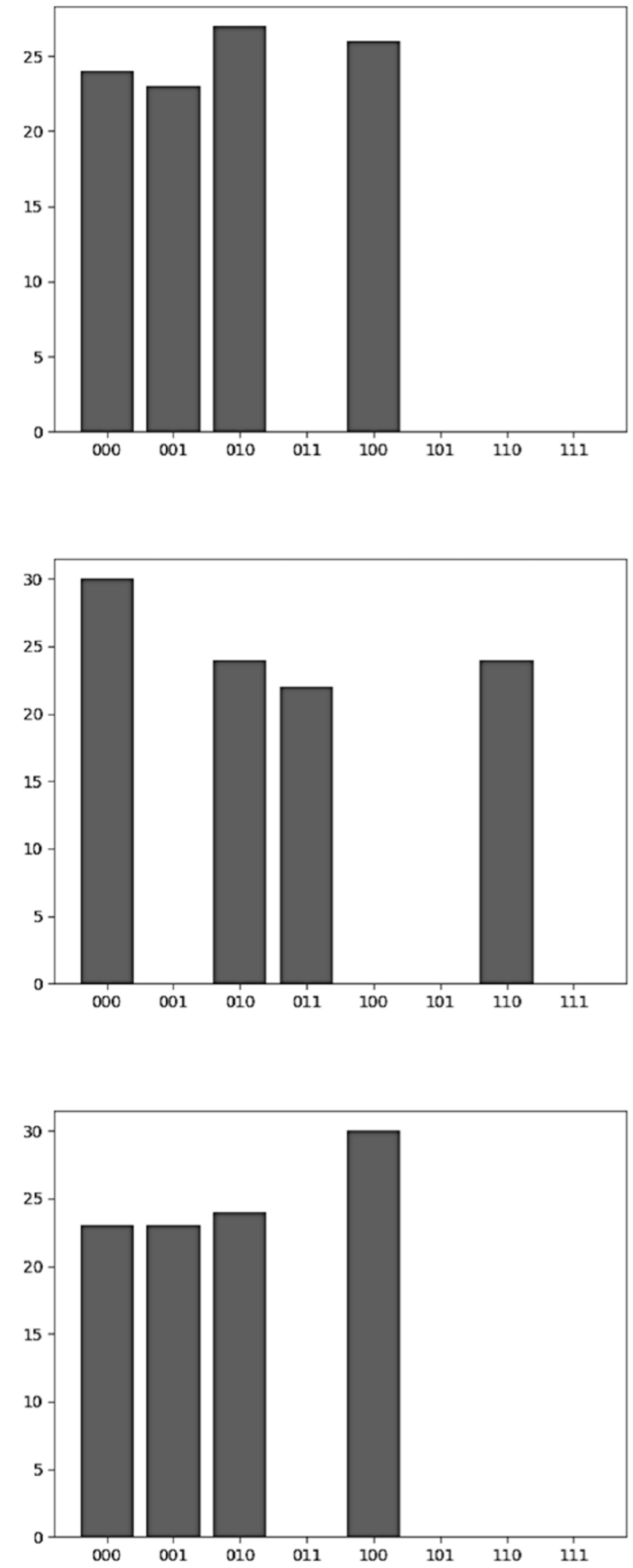

**Figure 27:** Histograms from 3 steps of the quantum walk algorithms for 100 shots each. For each step, the system selects the result that occurred more frequently: 010, 000 and 100, respectively.

Due to the statistical nature of quantum computation, it is often necessary to execute a quantum algorithm multiple times in order to obtain results that are statistically plausible. This enables one to inspect if the outcomes reflect the envisaged amplitude distribution of the quantum states. And running a circuit multiple times mitigates the effect of errors caused by undesired decoherence.

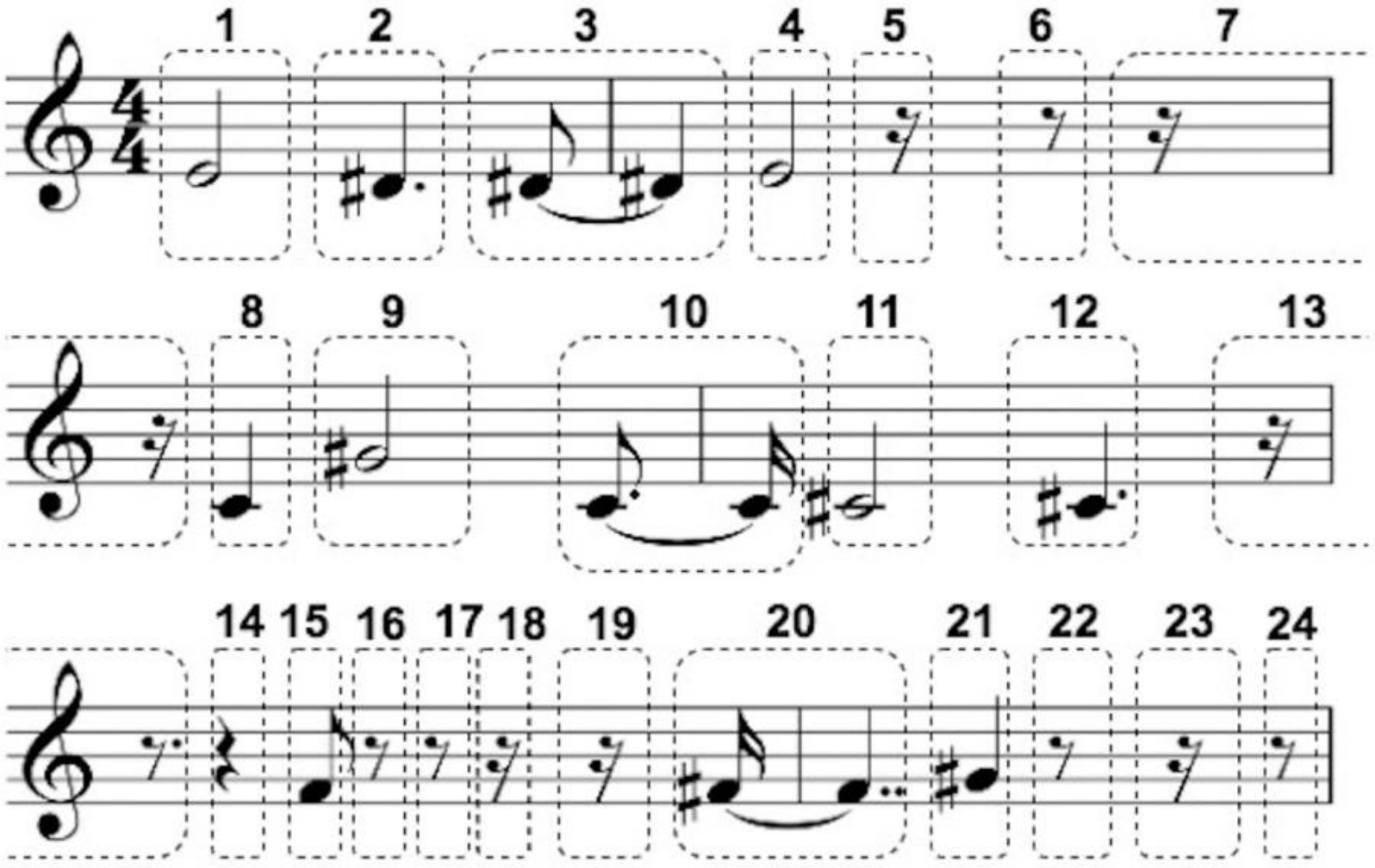

**Figure 28:** A music sequence generated by the random walk system.

| Step | Pitch | Duration |
|---|---|---|
| 1 | 110 | 100 |
| 2 | 010 | 110 |
| 3 | 010 | 110 |
| 4 | 110 | 100 |
| 5 | 100 | 101 |
| 6 | 100 | 001 |
| 7 | 000 | 001 |
| 8 | 000 | 000 |
| 9 | 001 | 100 |
| 10 | 000 | 000 |
| 11 | 100 | 100 |
| 12 | 100 | 110 |
| 13 | 000 | 010 |
| 14 | 010 | 010 |
| 15 | 011 | 011 |
| 16 | 111 | 001 |
| 17 | 011 | 001 |
| 18 | 111 | 101 |
| 19 | 111 | 101 |
| 20 | 101 | 100 |
| 21 | 001 | 000 |
| 22 | 000 | 001 |
| 23 | 000 | 101 |
| 24 | 000 | 001 |

**Table 4:** Pitch and duration codes generated for each step of the random walk example.

In quantum computing, the times an algorithm is run is referred to as *shots***.** For each shot, the measurements are stored in standard digital memory, and in the case of our quantum walk algorithm, the result that occurred more frequently is selected. Figure 27 shows histograms from running three steps of the quantum walk algorithms for 50 shots for generating pitches. Starting on 000, then the walker moves to 010, goes back to 000 and then it goes to 100. In this case the generated pitched were: D#3, C3 and C#3.

An example of a music sequence generated by the system is shown in Figure 28. In this case the system ran for 24 steps, 500 shots each. The initial pitch was 110 and the initial duration was 100. Table 4 shows the codes generated at each step of the walk.

## 6 Concluding remarks

Admittedly, the two quantum systems introduced above could as well be implemented on standard digital computers. At this stage, we are not advocating any quantum advantage for musical applications. What we advocate, however, is that the music technology community should be quantum-ready for when quantum computing hardware becomes more sophisticated, widely available, and possibly advantageous for creativity and business. In the process of learning and experimenting with this new technology, novel approaches, creative ideas, and innovative applications are bound to emerge.

The method introduced above to control the vocal synthesiser certainly begs further development. The codes to retrieve synthesis parameter values were assembled with three bits taken from a string of nine measurements. The algorithm to assemble the codes is as arbitrary as the association of codes to specific parameters; e.g., what is the rationale of allocating ($c_8$ $c_7$ $c_6$) to retrieve `fq1s`? Or why ($c_8$ $c_7$ $c_6$) instead of ($c_0$ $c_1$ $c_2$) or perhaps ($c_2$ $c_2$ $c_2$)?

Research is needed in order to forge stronger couplings between quantum computational processes and the synthesis parameters. Quantum computing should be used here to generate codes that are meaningful; it should be harnessed to yield values for producing targeted vocalizations; e.g., to sing syllables or words.

The complete vocal synthesiser shown above requires 52 parameter values to produce a sound. Moreover, the linear functions in fact need to be piecewise linear functions with various breakpoints in order to simulate transitions of vocal articulations. Therefore, the synthesis parameters' search space to produce a desired sung utterance is vast. It is here that quantum search algorithms may provide an advantageous solution [24] in the near future.

The quantum walk sequencer is an example of a first attempt at designing quantum versions of classic algorithmic music composition techniques. The shortcoming is that the sequencer has a limited number of musical parameters to work with; e.g., only eight notes. A larger number of parameters would require a much larger quantum circuit. But this increases the problem of decoherence, as mentioned earlier. Improved quantum hardware and better error correction methods will enable circuits with greater number qubits and gates in the future. In the meantime, simulators are available for research [1, 2].

It has been argued that quantum walk (on real quantum hardware) is faster than classical random walk to navigate vast mathematical spaces [25]. Quantum walk is an area of much interest for computer music. In addition to its generative uses, quantum walk is applicable as a search algorithm and in machine learning [26].

Musicians started experimenting with computers very early on in the history of computing, and paved the way for today's thriving music industry. CSIR's Mk1 was one of only a handful of electronic computers in existence at the time. And the mainframe used to compose the *Illiac Suit*e string quartet was one of the first computers built in a university in the USA, comprising thousands of vacuum tubes.

It is often said that today's quantum computers are in a development stage comparable to those clunky mainframes built in the mid of the last century. Time is ripe for musicians to embrace this emerging technology.

**Acknowledgement**

Many thanks to Bob Coecke and Konstantinos Meichanetzidis at the Department of Computer Science of the University of Oxford for inspiring discussions and advice. The author is immensely thankful to Rigetti Computing for continuing access to quantum hardware resources and technical support.